\documentclass[aps,prb,twocolumn,showpacs,superscriptaddress]{revtex4}
\usepackage{graphicx}
\usepackage{bm}
\usepackage{amsmath}
\newcommand{\ket}[1]{\ensuremath{|#1\rangle}}
\newcommand{\cre}[1]{\ensuremath{#1^{\, \dagger}}}
\newcommand{\ann}[1]{\ensuremath{#1^{\,}}}
\newcommand{\ave}[1]{\ensuremath{\langle#1\rangle}}
\renewcommand{\Im}{\mbox{Im}}
\renewcommand{\Re}{\mbox{Re}}

\begin{document}


\title{Interplay of Kondo and superconducting correlations in the nonequilibrium Andreev transport through a quantum dot}

\author{Yasuhiro Yamada}
\affiliation{Department of Physics, Kyoto University, Kyoto 606-8502, Japan}%
\author{Yoichi Tanaka}
\affiliation{Condensed Matter Theory Laboratory, RIKEN, Wako, Saitama 351-0198, Japan}%
\author{Norio Kawakami}
\affiliation{Department of Physics, Kyoto University, Kyoto 606-8502, Japan}%

\date{\today}

\begin{abstract}
Using the modified perturbation theory, we theoretically study the nonequilibrium Andreev transport through a quantum dot coupled to normal and superconducting leads (N-QD-S), which is strongly influenced by the Kondo and superconducting correlations. From the numerical calculation, we find that the renormalized couplings between the leads and the dot in the equilibrium states characterize the peak formation in the nonequilibrium differential conductance. In particular, in the Kondo regime, the enhancement of the Andreev transport via a Kondo resonance occurs in the differential conductance at a finite bias voltage, leading to an anomalous peak whose position is given by the renormalized parameters. In addition to the peak, we show that the energy levels of the Andreev bound states  give rise to other peaks in the differential conductance in the strongly correlated N-QD-S system. All these features of the nonequilibrium transport are consistent with those in the recent experimental results [R. S. Deacon {\it et al.}, Phys. Rev. Lett. {\bf 104}, 076805 (2010); Phys. Rev. B {\bf 81}, 12308 (2010)]. We also find that the interplay of the Kondo and superconducting correlations induces an intriguing pinning effect of the Andreev resonances to the Fermi level  and its counter position.
\end{abstract}


\maketitle

\section{Introduction}

Electron transport through nanofabrications has attracted much attention in the studies of fundamental quantum physics as well as potential future devices. In particular, a quantum dot (QD), which has discrete energy levels where the electrons are correlated, provides an ideal arena to study the local Coulomb interaction effect on the transport~\cite{KouwenhovenLP01fqd, AleinerIL02qei, ReimannSM02eso}. The magnetic doublet states with spin $1/2$ are stabilized at an isolated QD with an odd number of electrons and the strong Coulomb interaction, which results in the Coulomb blockade for the transport through the QD coupled to leads. At sufficiently low temperatures, however, the local moment of the doublet states is screened by the electrons of the leads owing to the Kondo effect, and thus the Kondo singlet is stabilized, resulting in an anomalous enhancement of the zero bias conductance~\cite{GlazmanLI88rkt, NgTK88ocr, Goldhaber-GordonD98kei, CronenwettSM98atk, vanderWielWG00tke, NygardJ00kpi}.

If we replace the leads by s-wave superconductors, a different situation arises; the doublet is not screened due to the lack of low-lying energy states of the leads. Even in the system, a singlet state can be stabilized due to the superconducting proximity effect, and the system thus shows a transition between the doublet and the singlet~\cite{GlazmanLI89rjc, ClerkAA00lop, VecinoE03jct, ChoiM-S04kea, SianoF04jct, OguriA04qpt, TanakaY07kei, MengT09sdo}. Away from the transition point, one of the two states becomes the ground state and the other an excited state which is localized at the QD, i.e., the Andreev bound state. In this system, however, it is difficult to directly observe the Andreev bound states via transport measurements because of a supercurrent and a multiple Andreev reflection process~\cite{BuitelaarMR02qdi, BuitelaarMR03mar, Jarillo-HerreroP06qst, vanDamJordenA06sri, EichlerA07eei, Sand-JespersenT07kat, BuizertC07kus, Grove-RasmussenK07kre, EichlerA09ttj, Grove-RasmussenK09sbs, KanaiY10eco}
.

Recently, Deacon {et. al.} have observed the Andreev bound states experimentally  not in the above system but in the system with a QD coupled to normal and superconducting leads (N-QD-S) where an Andreev reflection dominates the transport~\cite{DeaconRS10tso, DeaconRS10kat}. In the N-QD-S system, however, the doublet states should be replaced by the Kondo singlet state owing to the screening by the electrons of the normal lead (N-lead), leading to a crossover between the Kondo singlet to the superconducting singlet. A lot of studies have thus far focused on how the competition between the Kondo and superconducting correlations affects the Andreev transport experimentally~\cite{GraberMR04qdc, GraberMR04kri,DeaconRS10tso, DeaconRS10kat, HofstetterL10fpe} and theoretically~\cite{FazioR98rat, SchwabP99ati, ChoSY99pqt, ClerkAA00asa, CuevasJC01kei, SunQ-f01ekr, AvishaiY01tta, AonoT03qdi, KrawiecM04ett, SplettstoesserJ07pta, TanakaY07nrg, DomanskiT08ibp, DomanskiT08mtf,  YamadaY10eat, KoertingV10ntv}. Indeed, Kondo-type anomalous phenomena have been observed in the measurement of zero bias conductance in the recent experiment~\cite{DeaconRS10kat}.

Experimentally, characteristics of the Andreev bound states emerge under nonequilibrium steady-state conditions where a finite bias voltage is applied to the N-lead. Some theoretical studies have dealt with the nonequilibrium transport properties in an N-QD-S system with emphasis on the influence of the Kondo effect~\cite{FazioR98rat, SchwabP99ati, ChoSY99pqt, SunQ-f01ekr, AvishaiY01tta, AonoT03qdi, KrawiecM04ett, DomanskiT08ibp, DomanskiT08mtf, YamadaY10eat} and also on the Andreev bound states~\cite{KoertingV10ntv}. However, the coexistence of the phenomena related to the Kondo effect and the Andreev bound states in the experiments indicates the necessity of further theoretical studies; it is needed for the comprehensive understanding of the transport to include the Andreev bound state as well as the interplay between the Kondo and superconducting correlations into the theory. 

In this paper, we study the nonequilibrium Andreev transport, by taking into account the above different aspects of the N-QD-S system in a unified way. To this end, we employ the modified second order perturbation theory (MPT) used previously by Cuevas et al.~\cite{CuevasJC01kei} and extend it to the nonequilibrium steady-state conditions. The MPT was originally formulated in the equilibrium Anderson model~\cite{Martin-RoderoA82ans}, then has been used in several different systems, e.g., a quantum dot coupled to normal leads~\cite{YeyatiAL93ecr, TakagiO99mfe, AligiaAA06nmt, Martin-RoderoA08iaf, YeyatiAL99tim} and as a impurity solver for the dynamical mean-field theory~\cite{KajueterH96nip, PotthoffM97iso}. Furthermore, we exploit the exact solution of the QD-S system with an infinitely large superconducting gap, which still has the essence of the Andreev bound states, to improve the perturbation theory. By systematically examining the nonequilibrium transport properties in a wide variety of the system parameters, we demonstrate that the theoretical results obtained in this paper are qualitatively in agreement of the recent experiments.~\cite{DeaconRS10tso, DeaconRS10kat}
We note that a part of the present results was briefly reported in ref.~\onlinecite{YamadaY10eat}.

This paper is organized as follows. In Sec. II, the model Hamiltonian is introduced and we formulate the modified second-order perturbation theory in Keldysh-Nambu space of the Green's function. Section III, we assess the validity of our method in the equilibrium case and define the renormalized parameters which characterize the electron transport in the nonequilibrium states. The results of nonequilibrium transport are shown in Sec. IV. We also analyze the superconducting pair amplitude and the local density of states at the QD in the nonequilibrium states. The correspondence between the theoretical and experimental results is also discussed in this section. A summary is given in Sec. V.

\section{Model and Method}

\subsection{Model Hamiltonian and Keldysh Green's function in Nambu space}

In order to describe the electron transport in the N-QD-S system, we use a single level QD coupled to a normal metal and a superconductor, which is applicable for the system with large level spacing of the QD,
\begin{equation}
H = H_\mathrm{QD} + H_\mathrm{N} + H_\mathrm{S} + H_\mathrm{TN} + H_\mathrm{TS} \label{Hami},
\end{equation}
where 
\begin{eqnarray}
H_\mathrm{QD}&\!\!=\!\!&\epsilon_\mathrm{d}\sum_{\sigma} n_{d\sigma} + Un_{d\uparrow}n_{d\downarrow}, \\
H_\mathrm{N} &\!\!=\!\!& \sum_{k\sigma} (\epsilon_{k}^{N}-\mu_{N}) c_{k\sigma}^{\dagger}c_{k\sigma}^{},\\
H_\mathrm{S} &\!\!=\!\!& \sum_{q\sigma} (\epsilon_{q}^{S}-\mu_{S}) a_{q\sigma}^{\dagger}a_{q\sigma}^{} \!\!+\!\! \sum_{q}(\Delta_{S}a_{q\downarrow}^{\dagger}a_{-q\uparrow}^{\dagger} \!\!+\!\! \mathrm{H.c.}), \\
H_\mathrm{TN} &\!\!=\!\!& \sum_{k\sigma}(t_\mathrm{N}c_{k\sigma}^{\dagger}d_{\sigma}^{} + \mathrm{H.c.}), \\
H_\mathrm{TS} &\!\!=\!\!& \sum_{q\sigma}(t_\mathrm{S}a_{q\sigma}^{\dagger}d_{\sigma}^{} + \mathrm{H.c.}).
\end{eqnarray}
Here, $d_{\sigma}^{\dagger}$ creates an electron with spin $\sigma$ at the QD which has an energy level $\epsilon_{d}$ and the Coulomb interaction $U$. Here, $n_{d\sigma}\equiv d_{\sigma}^{\dagger}d_{\sigma}$. $c_{k\sigma}^{\dagger}$ ($a_{q\sigma}^{\dagger}$) denotes the creation operator of an electron with spin $\sigma$ and wave vector $k$ ($q$) in the normal (superconducting) lead. The superconducting lead is assumed to be described by the BCS Hamiltonian with a superconducting gap $\Delta_{S}=\Delta \exp(i\theta_{S})$. The QD is coupled to the normal and superconducting leads labeled by $\alpha=N, S$ with hybridization $t_{\alpha}$.

In order to define nonequilibrium steady states of the N-QD-S system with a bias voltage $V$, we treat the Coulomb interaction $U$ as a perturbation. The non-interacting problem can be solved exactly with the Keldysh Green's function technique in Nambu space, from which we can define the chemical potentials of the normal and superconducting leads as $\mu_{N}=eV$ and $\mu_{S}=0$, respectively. 

In the noninteracting case, several different Green's functions in Nambu space at the QD are defined as
\begin{eqnarray}
\bm{g}^{r} (t,t') &\!\!=\!\!& -i\theta(t-t')
\begin{pmatrix}
\ave{[\ann{d}_{\uparrow}(t), \cre{d}_{\uparrow}(t')]}_{0} & \ave{[\ann{d}_{\uparrow}(t), \ann{d}_{\downarrow}(t')]}_{0} \\
\ave{[\cre{d}_{\downarrow}(t), \cre{d}_{\uparrow}(t')]}_{0} & \ave{[\cre{d}_{\downarrow}(t), \ann{d}_{\downarrow}(t')]}_{0} \\
\end{pmatrix}, \\
\bm{g}^{a} (t,t') &\!\!=\!\!& i\theta(t'-t)
\begin{pmatrix}
\ave{[\ann{d}_{\uparrow}(t), \cre{d}_{\uparrow}(t')]}_{0} & \ave{[\ann{d}_{\uparrow}(t), \ann{d}_{\downarrow}(t')]}_{0} \\
\ave{[\cre{d}_{\downarrow}(t), \cre{d}_{\uparrow}(t')]}_{0} & \ave{[\cre{d}_{\downarrow}(t), \ann{d}_{\downarrow}(t')]}_{0} \\
\end{pmatrix}, \\
\bm{g}^{<} (t,t') &\!\!=\!\!& i
\begin{pmatrix}
\ave{\cre{d}_{\uparrow}(t')\ann{d}_{\uparrow}(t)}_{0} & \ave{\ann{d}_{\downarrow}(t')\ann{d}_{\uparrow}(t)}_{0} \\
\ave{\cre{d}_{\uparrow}(t')\cre{d}_{\downarrow}(t)}_{0} & \ave{\ann{d}_{\downarrow}(t')\cre{d}_{\downarrow}(t)}_{0} \\
\end{pmatrix}, \\
\bm{g}^{>} (t,t') &\!\!=\!\!& -i
\begin{pmatrix}
\ave{\ann{d}_{\uparrow}(t)\cre{d}_{\uparrow}(t')}_{0} & \ave{\ann{d}_{\uparrow}(t)\ann{d}_{\downarrow}(t')}_{0} \\
\ave{\cre{d}_{\downarrow}(t)\cre{d}_{\uparrow}(t')}_{0} & \ave{\cre{d}_{\downarrow}(t)\ann{d}_{\downarrow}(t')}_{0} \\
\end{pmatrix},
\end{eqnarray}
where $\bm{g}^{r}$ and $\bm{g}^{a}$ denote the retarded and advanced Green's functions, which are also used in the equilibrium case, and $\bm{g}^{<}$ and $\bm{g}^{>}$ represent the lesser and greater Green's functions. We consider a sufficiently wide band of electrons in the leads, in which the coupling strength $\Gamma_{N(S)} (\omega) \equiv \pi |t_{N(S)}| \sum_{k(q)} \delta(\omega-\epsilon_{k(q)}^{N(S)})$ becomes a constant $\Gamma_{N(S)}$. Integrating out the electron degrees of freedom in the two leads, we obtain the Fourier transformed Green's functions,
\begin{eqnarray}
\bm{g}^{r}(\omega)&\!\!=\!\!&\left( (\omega + i \eta) \bm{I}-\epsilon_{d}\bm{\sigma}_{3} - \bm{\Sigma}_{t}^{r}(\omega)\right)^{-1},\label{eq::g0r}\\
\bm{g}^{a}(\omega)&\!\!=\!\!&\left( (\omega - i \eta) \bm{I}-\epsilon_{d}\bm{\sigma}_{3} - \bm{\Sigma}_{t}^{a}(\omega)\right)^{-1},\label{eq::g0a}\\
\bm{g}^{<}(\omega)&\!\!=\!\!&-
\bm{g}^{r}(\omega)\bm{\Sigma}_{t}^{<}(\omega)\bm{g}^{a}(\omega),\label{eq::g0l} \\
\bm{g}^{>}(\omega)&\!\!=\!\!&-
\bm{g}^{r}(\omega)\bm{\Sigma}_{t}^{>}(\omega)\bm{g}^{a}(\omega),\label{eq::g0g}
\end{eqnarray}
where 
\begin{eqnarray}
\bm{\Sigma}_{t}^{r}(\omega)&\!\!=\!\!&
\begin{pmatrix}
-i\left(\Gamma_{N}+\Gamma_{S}\beta(\omega)\right)  & i\Gamma_{S}\beta(\omega)\frac{\Delta_{S}}{\omega} \\
i\Gamma_{S}\beta(\omega)\frac{\Delta_{S}^{*}}{\omega} & -i\left(\Gamma_{N}+\Gamma_{S}\beta(\omega)\right) 
\end{pmatrix}, \\
\bm{\Sigma}_{t}^{a}(\omega)&\!\!=\!\!&\left[\bm{\Sigma}_{t}^{r}(\omega)\right]^{\dagger}, \\
\bm{\Sigma}_{t}^{<}(\omega)&\!\!=\!\!&-i2\Gamma_{N}
\begin{pmatrix}
f(\omega-\mu_{N}) & 0 \\
0 & f(\omega+\mu_{N})
\end{pmatrix} \nonumber \\
&&-i2\Gamma_{S}\Re[\beta(\omega)]
\begin{pmatrix}
1 & -\frac{\Delta_{S}}{\omega} \\
-\frac{\Delta_{S}^{*}}{\omega} & 1 \\
\end{pmatrix}
f(\omega), \\
\bm{\Sigma}_{t}^{>}(\omega)&\!\!=\!\!&i2\Gamma_{N}
\begin{pmatrix}
1-f(\omega-\mu_{N}) & 0 \\
0 & 1-f(\omega+\mu_{N})
\end{pmatrix} \nonumber \\
&&+i2\Gamma_{S}\Re[\beta(\omega)]
\begin{pmatrix}
1 & -\frac{\Delta_{S}}{\omega} \\
-\frac{\Delta_{S}^{*}}{\omega} & 1 \\
\end{pmatrix} 
\left(1-f(\omega)\right).
\end{eqnarray}
Here, $\eta$ is a positive infinitesimal and $\bm{\sigma}_{i}$ ($i=1,2,3$) is a Pauli matrix in Nambu space. $\beta(\omega)=\frac{|\omega|}{\sqrt{\omega^2 - \Delta^{2}}}\theta\left(|\omega| - \Delta\right)+\frac{\omega}{i\sqrt{\Delta^{2}-\omega^2}}\theta\left(\Delta-|\omega|\right)$ and $f(x)=[e^{x/k_{B}T}+1]^{-1}$. 

If we obtain the self-energies due to the Coulomb interaction, the full retarded and advanced Green's functions are determined from the Dyson equation,
\begin{eqnarray}
\bm{G}^{r,a}(\omega)\!\!&=&\!\!\left[{\left[\bm{g}^{r,a}(\omega)\right]}^{-1}-\bm{\Sigma}_{U}^{r,a} (\omega) \right]^{-1}. \label{eq::Gra}
\end{eqnarray}
The full lesser and greater ones are calculated from the relation,
\begin{equation}
\bm{G}^{<,>}(\omega)=-\bm{G}^{r}(\omega)\Big[\bm{\Sigma}_{t}^{<,>}(\omega) + \bm{\Sigma}_{U}^{<,>}(\omega) \Big]\bm{G}^{a}(\omega).\label{eq::Glg}
\end{equation}

\begin{figure}
\includegraphics[width=8cm,clip]{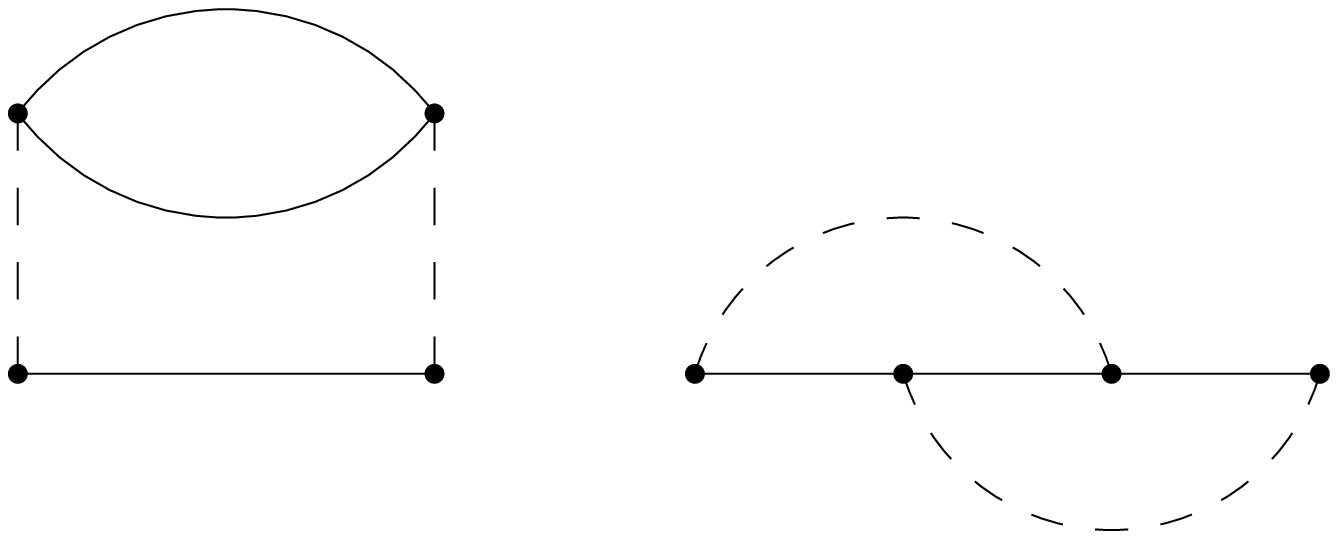}
\caption{Second order self-energy diagrams we consider in this paper. The solid line indicates the propagator and the dashed line the Coulomb interaction.}
\label{fig:2nd_diagram} 
\end{figure}
Here we examine the self-energy $\bm{\Sigma}_{U}$, using the perturbation theory in the Keldysh Green's function formalism. The first order contributions to the retarded and advanced self-energies are $\bm{\Sigma}_{\mathrm{1st}}^{r,a}=U\ave{\bm{N}}$ with
\begin{equation}
\ave{\bm{N}}\equiv
\begin{pmatrix}
\ave{n_{d}} & \ave{d_{\downarrow}d_{\uparrow}} \\
\ave{d_{\downarrow}d_{\uparrow}}^{*} & -\ave{n_{d}}
\end{pmatrix},
\end{equation}
where $\ave{n_{d}}$ denotes the expectation value of the electron number at the QD per spin. There is no first-order contribution to the lesser and greater self-energies because the Coulomb interaction at the QD takes place without delay. The second order skeleton diagrams are depicted in Fig. \ref{fig:2nd_diagram}. The corresponding contributions to the lesser and greater self-energies are obtained from the equations,
\begin{eqnarray}
\bm{\Sigma}_\mathrm{2nd}^{<}(\omega)=
-U^2\int{}\frac{d\omega_{1}}{2\pi}\Pi^{<}(\omega+\omega_{1})\bm{\sigma}_{2}\left(\bm{g}^{>}(\omega_{1})\right)^{T}\bm{\sigma}_{2}, \\
\bm{\Sigma}_\mathrm{2nd}^{>}(\omega)=
-U^2\int{}\frac{d\omega_{1}}{2\pi}\Pi^{>}(\omega+\omega_{1})\bm{\sigma}_{2}\left(\bm{g}^{<}(\omega_{1})\right)^{T}\bm{\sigma}_{2},
\end{eqnarray}
where
\begin{eqnarray}
\Pi^{<,>}(\omega)&\!\!=\!\!&\int\frac{d\omega_{1}}{2\pi}\Big(
g_{11}^{<,>}(\omega_{1})g_{22}^{<,>}(\omega-\omega_{1}) \nonumber\\
&&\qquad \qquad- g_{12}^{<,>}(\omega_{1})g_{21}^{<,>}(\omega-\omega_{1})\Big),
\end{eqnarray}
where $g_{ij}$ denotes the $(i ,j)$ component of $\bm{g}$. Using the above self-energies, we calculate the second order contributions to the retarded and advanced self-energies, 
\begin{equation}
\bm{\Sigma}_\mathrm{2nd}^{r,a}(\omega)=
\frac{i}{2\pi}\int\frac{d\omega_{1}}{\omega-\omega_{1} \pm i\eta}\left[\bm{\Sigma}_\mathrm{2nd}^{<}(\omega_{1})-\bm{\Sigma}_\mathrm{2nd}^{>}(\omega_{1})\right].\end{equation}

Although the second order self-energies are believed to give reasonable results for the nonequilibrium transport through a strongly interacting QD coupled to two normal leads (N-QD-N system) at least for the particle-hole symmetric case~\cite{HershfieldS91ptk, HershfieldS92rtt}, this technique is not directly applicable to the N-QD-S system because of the lack of symmetry of the leads. Main difficulty in our N-QD-S system comes from the fact that the simple second order self-energies do not correctly give the formula in the "atomic limit" where the QD and the leads are disconnected. Indeed, the qualitatively correct description of the Kondo effect in a particle-hole symmetric N-QD-N system with the second order self-energy is ensured by the fact that the corresponding formula becomes exact not only in the weak-$U$ but also in the atomic limit \cite{KajueterH96nip}. In the N-QD-S system, furthermore, the superconducting correlations at the QD, which come from the superconducting proximity effects, must be taken into account, so that we have to introduce a suitable "atomic limit"  to study the strong-$U$ regime. For this purpose, we here make use of the exact solution of the QD-S system with an infinitely large superconducting gap. 

Below, we describe how to construct the modified self-energies, $\widetilde{\bm{\Sigma}}_{\mathrm{2nd}}$, which reproduce the correct results in the atomic limit as well as the weak-$U$ limit within the above second-order perturbation. The self-energy due to the Coulomb interaction up to second order thus reads $\bm{\Sigma}_{U}=\bm{\Sigma}_{\mathrm{1st}} + \widetilde{\bm{\Sigma}}_{\mathrm{2nd}}$.

\subsection{Superconducting atomic limit}

In the limit of $\Delta \to \infty$, the quasiparticle degree of freedom in the superconducting lead is decoupled from the QD. Therefore, the Hamiltonian is simplified as,
\begin{eqnarray}
H^{\Delta\mathrm{inf}} &\!\!=\!\!& H_\mathrm{dot}^{\Delta\mathrm{inf}}+H_{N} + H_{TN}, \label{eq::HamiDinf}
\\
H_\mathrm{dot}^{\Delta\mathrm{inf}} &\!\!=\!\!& \epsilon_{d}\sum_{\sigma} n_{d\sigma} + (\Delta_{d}\cre{d_{\uparrow}}\cre{d_{\downarrow}} + \mathrm{H.c.}) + Un_{{d}\uparrow}n_{{d}\downarrow}, \label{eq::HamiDinfdot}
\end{eqnarray}
where $\Delta_{d}\equiv\Gamma_{S}\exp(i\theta_{S})$. In this case, $\Gamma_{S}$ corresponds to the effective superconducting gap at the QD owing to the proximity effect.

In the limit of $\Gamma_{N} \to 0$, the QD is decoupled from the normal lead and the effective Hamiltonian(\ref{eq::HamiDinf}) becomes a one-site problem with the superconducting paring potential and the Coulomb interaction. Hereafter, we call the limit of ($\Delta \to \infty, \Gamma_{N}/U \to 0$) "superconducting atomic limit"~\cite{MengT09sdo}. In the superconducting atomic limit, the self-energies at the QD can be exactly obtained as,
\begin{eqnarray}
\bm{\Sigma}_{\textrm{atm}}^{r}(\omega)&\!\!=\!\!&U^2 \chi \left((\omega+i\eta)\bm{I}-\bm{\Delta}\right)^{-1}, \label{eq::ratmlimit} \\
\bm{\Sigma}_{\textrm{atm}}^{a}(\omega)&\!\!=\!\!&U^2 \chi \left((\omega-i\eta)\bm{I}-\bm{\Delta}\right)^{-1}, \label{eq::aatmlimit} \\
\bm{\Sigma}_{\textrm{atm}}^{<}(\omega)&\!\!=\!\!& -i2\eta U^2 \chi \left((\omega+i\eta)\bm{I}-\bm{\Delta}\right)^{-1} \nonumber \\
&&\: \times ((\bm{I}-\bm{\sigma}_{3})/2 +\ave{\bm{N}})\nonumber \\
&&\: \times \left((\omega-i\eta)\bm{I}-\bm{\Delta}\right)^{-1}, \label{eq::gatmlimit} \\
\bm{\Sigma}_{\textrm{atm}}^{>}(\omega)&\!\!=\!\!& i2\eta U^2 \chi \left((\omega+i\eta)\bm{I}-\bm{\Delta}\right)^{-1} \nonumber \\
&&\: \times ((\bm{I}+\bm{\sigma}_{3})/2-\ave{\bm{N}}) \nonumber \\
&&\: \times \left((\omega-i\eta)\bm{I}-\bm{\Delta}\right)^{-1}, \label{eq::latmlimit}
\end{eqnarray}
where
\begin{eqnarray}
\chi&\equiv& \langle n_{d} \rangle (1- \langle n_{d} \rangle )-| \langle d_{\downarrow}d_{\uparrow} \rangle |^{2}, \\
\bm{\Delta}&\equiv&
\begin{pmatrix}
\epsilon_{d}+U(1- \langle n_{d} \rangle )&
\Delta_{d} - U\ave{d_{\downarrow}d_{\uparrow}} \\
\Delta_{d}^{*} - U\ave{d_{\downarrow}d_{\uparrow}}^{*} & -\epsilon_{d}-U(1-\ave{n_{d}}) 
\end{pmatrix}.
\end{eqnarray}
Note that we omit the first order contributions of $U$ in the atomic-limit self-energies.

Next, we consider the second order self-energies, following the formula derived in the previous section. We assume that one-particle Green's functions in the second order diagrams (Fig. \ref{fig:2nd_diagram}) are dressed with energy shifts, which are determined by the following one-body Hamiltonian,
\begin{eqnarray}
\overline{H} &\!\!=\!\!& \overline{H}_\mathrm{dot}+H_{N} + H_{S} + H_{TN} + H_{TS}, \label{eq::onebodyHami} \\
\overline{H}_\mathrm{dot} &=& \left(\epsilon_{d}+U\ave{\overline{n_{d}}}\right) \sum_{\sigma} \cre{d_{\sigma}} \ann{d_{\sigma}} \nonumber \\
&&\quad+(U\ave{\overline{d_{\downarrow}d_{\uparrow}}}\cre{d_{\uparrow}}\cre{d_{\downarrow}} + H.c.),
\end{eqnarray}
where $\ave{\overline{n_{d}}}$ and $\ave{\overline{d_{\downarrow}d_{\uparrow}}}$ are the effective parameters representing the energy shifts of the one-particle Green's function.

In the superconducting atomic limit, the above one-particle Green's functions behave like a $\delta$ function, so that the second order self-energies can be evaluated as,
\begin{eqnarray}
\bm{\Sigma}_{\textrm{2nd}}^{r}(\omega)&\!\!\to\!\!&U^2 \chi_{0} \left((\omega+i\eta)\bm{I}-\bm{\Delta}_{0}\right)^{-1}, \label{eq::r2ndlimit}\\
\bm{\Sigma}_{\textrm{2nd}}^{a}(\omega)&\!\!\to\!\!&U^2 \chi_{0} \left((\omega-i\eta)\bm{I}-\bm{\Delta}_{0}\right)^{-1}, \label{eq::a2ndlimit}\\
\bm{\Sigma}_{\textrm{2nd}}^{<}(\omega)&\!\!\to\!\!& -i2\eta U^2 \chi_{0} \left((\omega+i\eta)\bm{I}-\bm{\Delta}_{0}\right)^{-1} \nonumber \\
&& \times ((\bm{I}-\bm{\sigma}_{3})/2+\ave{\bm{N}}_{0}) \nonumber \\ 
&& \times \left((\omega-i\eta)\bm{I}-\bm{\Delta}_{0}\right)^{-1}, \label{eq::l2ndlimit}\\
\bm{\Sigma}_{\textrm{2nd}}^{>}(\omega)&\!\!\to\!\!& i2\eta U^2 \chi_{0} \left((\omega+i\eta)\bm{I}-\bm{\Delta}_{0}\right)^{-1}\nonumber \\ 
&& \times ((\bm{I}+\bm{\sigma}_{3})/2-\ave{\bm{N}}_{0})\nonumber \\
&& \times \left((\omega-i\eta)\bm{I}-\bm{\Delta}_{0}\right)^{-1}, \label{eq::g2ndlimit}
\end{eqnarray}
where
\begin{eqnarray}
\chi_{0}&\equiv&\ave{n_{d}}_{0}(1-\ave{n_{d}}_{0})-|\ave{d_{\downarrow}d_{\uparrow}}_{0}|^{2},\\
\bm{\Delta}_{0}&\equiv&
\begin{pmatrix}
\epsilon_{d}+U\ave{\overline{n_{d}}} & \Delta_{d} + U\ave{\overline{d_{\downarrow}d_{\uparrow}}}\\
\Delta_{d}^{*} + U\ave{\overline{d_{\downarrow}d_{\uparrow}}}^{*} & -\epsilon_{d}-U\ave{\overline{n_{d}}}
\end{pmatrix} ,\\
\ave{\bm{N}}_{0}&\equiv&
\begin{pmatrix}
\ave{n_{d}}_{0} & \ave{d_{\downarrow}d_{\uparrow}}_{0} \\
\ave{d_{\downarrow}d_{\uparrow}}_{0}^{*} & -\ave{n_{d}}_{0}
\end{pmatrix}.
\end{eqnarray}
Here, $\ave{n_{d}}_{0}$ and $\ave{d_{\downarrow}d_{\uparrow}}_{0}$ are the expectation values of the particle number per spin and the superconducting correlation at the QD under the one-body Hamiltonian \eqref{eq::onebodyHami}.

We find that in the superconducting atomic limit, the second-order self-energies have the functional forms similar to the exact ones, except that they have different constants; ($\chi_{0}$, $\bm{\Delta}_{0}$, $\ave{\bm{N}}_{0}$) and ($\chi$, $\bm{\Delta}$, $\ave{\bm{N}}$). Exploiting this fact, we construct the modified second order self-energies in the following. 

\subsection{Modified second order perturbation theory}

First, we formulate the modified self-energies for the retarded and advanced sectors. For the sake of clarity, we follow the procedure of the modified perturbation theory (MPT) in the N-QD-N system by Kajueter and Kotliar \cite{KajueterH96nip}. In this procedure, the modified self-energies are assumed to have the following functional forms,
\begin{equation}
\left\{
\begin{aligned}
\bm{\widetilde{\Sigma}}_{\mathrm{2nd}}^{\,r}(\omega)&=A
\left[\left[\bm{\Sigma}_{\mathrm{2nd}}^{\,r}(\omega)\right]^{-1}-\bm{B}\right]^{-1}\\
\bm{\widetilde{\Sigma}}_{\mathrm{2nd}}^{\,a}(\omega)&=A
\left[\left[\bm{\Sigma}_{\mathrm{2nd}}^{\,a}(\omega)\right]^{-1}-\bm{B}\right]^{-1}
\end{aligned}
\right. \label{eq::ra2ndmod}
\end{equation}
where $A$ and $\bm{B}$ should be determined for the self-energies to reproduce the exact ones in the high-energy limit as well as the superconducting atomic limit. The exact high-energy limit of the self-energies can be calculated from the continued-fraction expansion of the corresponding Green's functions \cite{GordonRG68ebi, NoltingW89bmi},
\begin{eqnarray}
\bm{\Sigma}_{U}^{r,a}(\omega)=U\ave{\bm{N}}
+ \frac{U^2\chi}{\omega} \bm{I} +O(\frac{1}{\omega^2}). \label{eq::exacthighenergy}
\end{eqnarray}
The first term coincides with the first order self-energy in $U$. 
On the other hand, in the limit of $\omega\to\infty$, the modified self-energies are expanded as,
\begin{eqnarray}
\widetilde{\bm{\Sigma}}_{\textrm{2nd}}^{r,a}(\omega)=
\frac{AU^2\chi_{0}}{\omega} \bm{I} +O(\frac{1}{\omega^2}). \label{eq::2ndhighenergy}
\end{eqnarray}
The coefficient $A$ in eq. \eqref{eq::2ndhighenergy} is determined from the condition that the leading terms of the modified self-energies are identical to the corresponding ones in the exact self-energies in eq. \eqref{eq::exacthighenergy}. Accordingly, we set $A$ as $\chi/\chi_{0}$.

 We next determine the matrix $\bm{B}$ from the condition that the modified self-energies give the correct values in the superconducting atomic limit. In the limit, the modified retarded and advanced self-energies become
\begin{equation}
\bm{\widetilde{\Sigma}}_{\mathrm{2nd}}^{r, a}(\omega)\to U^2\chi ((\omega\pm i\eta)\bm{I}-\bm{\Delta}_{0}-U^2\chi_{0} \bm{B})^{-1}.
\end{equation}
In order to eliminate the difference between the r.h.s of the above equation and that of eqs. \eqref{eq::ratmlimit} and \eqref{eq::aatmlimit}, we set $\bm{B}$ as follows,
\begin{equation}
\bm{B}=\frac{1}{U \chi_{0}}\begin{pmatrix}
1-\ave{n_{d}}-\ave{\overline{n_{d}}} & -\ave{d_{\downarrow}d_{\uparrow}}-\ave{\overline{d_{\downarrow}d_{\uparrow}}} \\ 
-\ave{d_{\downarrow}d_{\uparrow}}^{*}-\ave{\overline{d_{\downarrow}d_{\uparrow}}}^{*} & -1+\ave{n_{d}}+\ave{\overline{n_{d}}}
\end{pmatrix}.
\end{equation}
Note that in the limit of $\Delta \to 0$ or $\Gamma_{S} \to 0$, the superconducting correlations at the QD vanishes and the off-diagonal terms in eq. \eqref{eq::ra2ndmod} become zero. Furthermore, $A=\ave{n_{d}}\left(1-\ave{n_{d}}\right)/(\ave{n_{d}}_{0}\left(1-\ave{n_{d}}_{0}\right))$ and the non-diagonal terms of $\bm{B}$ vanish. As a result, the modified self-energies of eq.\eqref{eq::ra2ndmod} just coincide with those in the previous studies\cite{KajueterH96nip}. Therefore,  we believe that the modified self-energies obtained here are proper extensions of those used in the N-QD-N system.

We have to calculate the modified lesser and greater self-energies in order to obtain the transport properties. By generalizing the strategy used previously\cite{KajueterH96nip}, we define the modified lesser self-energy as,
\begin{eqnarray}
\widetilde{\bm{\Sigma}}_{\textrm{2nd}}^{<}(\omega)&=&\frac{1}{A}
\widetilde{\bm{\Sigma}}_{\textrm{2nd}}^{r}(\omega)
\left[\bm{\Sigma}_{\textrm{2nd}}^{\,r}(\omega)\right]^{-1} \bm{\Sigma}_{\textrm{2nd}}^{<}(\omega)\nonumber \\
&& \quad \times \left[\bm{\Sigma}_{\textrm{2nd}}^{\,a}(\omega)\right]^{-1}
\widetilde{\bm{\Sigma}}_{\textrm{2nd}}^{a}(\omega) \label{eq::l2ndmod}
\end{eqnarray}
We now confirm that the above self-energy reproduces the atomic-limit form. Multiplying the modified self-energy (\ref{eq::l2ndmod}) on the left and right by the inverse matrices of $\widetilde{\bm{\Sigma}}_{\textrm{2nd}}^{r}(\omega)$ and $\widetilde{\bm{\Sigma}}_{\textrm{2nd}}^{a}(\omega)$, we take the superconducting atomic limit,
\begin{eqnarray}
&&\left[\widetilde{\bm{\Sigma}}_{\textrm{2nd}}^{r}(\omega)\right]^{-1}
\widetilde{\bm{\Sigma}}_{\textrm{2nd}}^{<}(\omega) 
\left[\widetilde{\bm{\Sigma}}_{\textrm{2nd}}^{a}(\omega)\right]^{-1} \nonumber\\ 
&&\to \frac{-2i\eta}{U^2 \chi}((\bm{I}-\bm{\sigma}_{3})/2+\ave{\bm{N}}_{0}). \label{eq::ml2ndlimitm}
\end{eqnarray}
In the limit, the above  matrix does not depend on $\omega$. In a similar way, we multiply the atomic-limit lesser self-energy (\ref{eq::l2ndmod}) on the left and right by the same matrices. The resulting matrix also becomes a constant in the limit,
\begin{eqnarray}
&&\left[\widetilde{\bm{\Sigma}}_{\textrm{2nd}}^{r}(\omega)\right]^{-1}
\bm{\Sigma}_{\textrm{atm}}^{<}(\omega) 
\left[\widetilde{\bm{\Sigma}}_{\textrm{2nd}}^{a}(\omega)\right]^{-1} \nonumber\\ &&\to \frac{-2i\eta}{U^2 \chi}((\bm{I}-\bm{\sigma}_{3})/2+\ave{\bm{N}}). \label{eq::mlatmlimitm}
\end{eqnarray}
Therefore, the difference between eq. (\ref{eq::ml2ndlimitm}) and eq. (\ref{eq::mlatmlimitm}) can be ignored in the limit of $\eta \to 0$, and eq. (\ref{eq::l2ndmod}) reproduces the atomic-limit form indeed,
\begin{eqnarray}
\widetilde{\bm{\Sigma}}_{\textrm{2nd}}^{<}(\omega) \to
\bm{\Sigma}_{\textrm{atm}}^{<}(\omega) 
 && (\Delta \to \infty, \Gamma_{N} \to 0).
\end{eqnarray}
We also define the modified greater self-energy as,
\begin{eqnarray}
\widetilde{\bm{\Sigma}}_{\textrm{2nd}}^{>}(\omega)&=&\frac{1}{A}
\widetilde{\bm{\Sigma}}_{\textrm{2nd}}^{r}(\omega)
\left[\bm{\Sigma}_{\textrm{2nd}}^{\,r}(\omega)\right]^{-1} \bm{\Sigma}_{\textrm{2nd}}^{>}(\omega)\nonumber \\
&& \quad \times \left[\bm{\Sigma}_{\textrm{2nd}}^{\,a}(\omega)\right]^{-1}
\widetilde{\bm{\Sigma}}_{\textrm{2nd}}^{a}(\omega), \label{eq::g2ndmod}
\end{eqnarray}
which gives an appropriate form in the limit.

So far, we have formulated the modified retarded, advanced, lesser and greater self-energies. However, these four self-energies are not independent, but have to satisfy the following equality, 
\begin{eqnarray}
\widetilde{\bm{\Sigma}}_{\textrm{2nd}}^{<}(\omega) - \widetilde{\bm{\Sigma}}_{\textrm{2nd}}^{>}(\omega)&=\widetilde{\bm{\Sigma}}_{\textrm{2nd}}^{r}(\omega)-\widetilde{\bm{\Sigma}}_{\textrm{2nd}}^{a}(\omega).
\end{eqnarray}
We indeed confirm that the modified self-energies satisfy the equality.

\subsection{Current conservation and consistency of the energy shifts}

Using the resulting full Green's function, we calculate the current though the N-QD-S system. The current flowing in the normal (N) and superconducting (S) leads can be calculated from the time evolution of the particle number operators   $\hat{N}_{N,S}$ in each lead: $\hat{I}_{N}(t)=-e\frac{\mathrm{d} \hat{N}_{N}(t)}{\mathrm{d}t}$ and $\hat{I}_{S}(t)=-e\frac{\mathrm{d} \hat{N}_{S}(t)}{\mathrm{d}t}$. Since we assume that the system is in a nonequilibrium steady state, the expectation values of these operators are time-independent, which are given by 
\begin{equation}
\langle \hat{I}_{N} \rangle =-\frac{4e\Gamma_{N}}{h}\Im\int\left[2 f(\omega-\mu_{N})G_{11}^{r}(\omega)+G_{11}^{<}(\omega)\right] d \omega,\label{eq::currentN}
\end{equation}
\begin{eqnarray}
\!\!\!\!\!\!\langle \hat{I}_{S} \rangle &\!\!=\!\!&-\frac{4e\Gamma_{S}}{h}\Im\int{d}\omega \nonumber \\
\!\!\!\!\!\!&\!\!\!\!&\times\Bigg[2\tilde{\rho}_{S}(\omega)f(\omega)G_{11}^{r}(\omega)+\beta^{*}(\omega)G_{11}^{<}(\omega) \nonumber \\
\!\!\!\!\!\!&\!\!\!\!&-\frac{\Delta_{S}}{\omega}\bigg(2\tilde{\rho}_{S}(\omega)f(\omega)G_{12}^{r}(\omega)+\beta^{*}(\omega)G_{12}^{<}(\omega)\bigg) \Bigg], \label{eq::currentS}
\end{eqnarray}
where $\tilde{\rho}_{S}(\omega) \equiv \Re[\beta(\omega)]$ and $G_{ij}$ denotes the $(i, j)$ component of $\bm{G}$. We define the current  $I$ in this system as $I=\langle \hat{I}_{N} \rangle =- \langle \hat{I}_{S} \rangle $. However, it is known that the current calculated by the second order perturbation theory may not be conserved in some quantum dot systems except in a special condition \cite{HershfieldS91ptk, HershfieldS92rtt, Dell'AnnaL08snt}. In the N-QD-S system, the simple application of the second order self-energy usually breaks the current conservation rule, i.e. $\langle \hat{I}_{N} \rangle + \langle \hat{I}_{S} \rangle \neq 0$.

In our modified second order perturbation theory, the problem in the current still exists. In order to resolve this difficulty within our framework, we introduce the source term $\lambda$ coupled to the current operator and add the term into the one-body Hamiltonian \eqref{eq::onebodyHami},

\begin{equation}
\lambda(I_{N}+I_{S}).
\end{equation}

The effective parameters $\ave{\overline{n_{d}}}$, $\ave{\overline{d_{\downarrow}d_{\uparrow}}}$ and $\lambda$ are determined by the following consistency  conditions on the energy shifts and the current conservation,
\begin{equation}
\left\{
\begin{aligned}
U\ave{\overline{n_{d}}}&= U\ave{n_{d}}+\Re[\widetilde{\bm{\Sigma}}^{r}_{\mathrm{2nd}}(\mu_{N})]_{11} \\
U\ave{\overline{d_{\downarrow}d_{\uparrow}}}&= U\ave{d_{\downarrow}d_{\uparrow}}+[\widetilde{\bm{\Sigma}}^{r}_{\mathrm{2nd}}(\mu_{S})]_{12} \\
\langle \hat{I}_{N} \rangle + \langle \hat{I}_{S} \rangle &= 0,
\end{aligned}
\right. \label{eq::conditions}
\end{equation}
where $[\widetilde{\bm{\Sigma}}^{r}_{\mathrm{2nd}}]_{ij}$ denotes the $(i,j)$ component of the modified retarded self-energy. $\ave{n_{d}}$ and $\ave{d_{\downarrow}d_{\uparrow}}$ are also determined in a self-consistent manner. 

Here, we check the $U \to 0$ limit of the modified self-energies. In the small-$U$ limit,  $\widetilde{\bm{\Sigma}}^{r,a}_{\mathrm{2nd}}$ becomes the simple second order self-energy $\bm{\Sigma}^{r,a}_{\mathrm{2nd}}$ which is calculated from the one-particle Green's functions dressed with the mean-field energy shift because the consistency conditions in eq \eqref{eq::conditions} are reduced to $\ave{\overline{n_{d}}}=\ave{n_{d}}=\ave{n_{d}}_{0} $ and $\ave{\overline{d_{\downarrow}d_{\uparrow}}}=\ave{d_{\downarrow}d_{\uparrow}}=\ave{d_{\downarrow}d_{\uparrow}}_{0}$. Therefore, $\widetilde{\bm{\Sigma}}^{<,>}_{\mathrm{2nd}}$ is also reduced to  $\bm{\Sigma}^{<,>}_{\mathrm{2nd}}$ evaluated with using the mean-field Green's functions.

Note that the above modified self-energies are applicable to impurity systems with or without superconducting correlation. We will show below that the above method works very well except for some special cases with large bias voltage where we cannot find the convergent parameters $\ave{\overline{n_{d}}}$, $\ave{\overline{d_{\downarrow}d_{\uparrow}}}$ and $\lambda$. In this paper, we mainly focus on the reasonable parameter region where the bias voltage is not so large. We demonstrate that a variety of intriguing phenomena emerge due to the interplay between the superconducting correlation and the Kondo effect, some of which indeed  reproduce the experimental results qualitatively well.

\section{ Linear-response conductance and phase diagram}
In this section, we study the transport properties in the linear-response regime and check the validity of our approximation for the electron transport. In addition, the renormalized couplings of tunneling are introduced, which clearly specify various regimes appearing in the nonequilibrium electron transport addressed in the next section. 

\subsection{Zero bias conductance and the renormalized couplings in the equilibrium states} 

Let us first consider the zero bias conductance obtained in two different ways within the same framework of modified perturbation theory (MPT) to confirm the consistency of our approximation. Here, we concentrate on the symmetric coupling case, $\Gamma_{N}/\Gamma_{S}=1$, with particle-hole symmetry, $\epsilon_{d}/U=-0.5$, in the equilibrium state.

\begin{figure}
\includegraphics[width=8cm,clip]{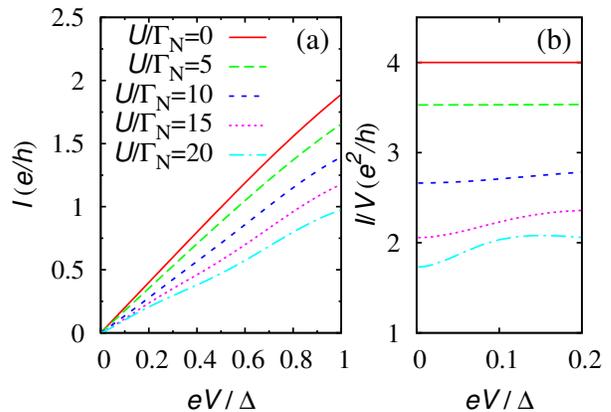}
\caption{\label{fig:current} (Color online) (a) Current-voltage characteristics for several values of $U$: $\Gamma_{S}/\Gamma_{N}=1$, $\epsilon_{d}/U=-0.5$, $\Delta/\Gamma_{N}=0.5$ and $k_{B}T/\Gamma_{N}=0.005$. (b) Conductance as a function of the bias voltage. The parameters used are the same as in (a).}
\end{figure}
We first obtain the zero bias conductance by directly differentiating the current by the bias voltage. In this case, we have to calculate the lesser and greater self-energies in order to obtain the current from eqs. \eqref{eq::currentN} and \eqref{eq::currentS}. The current-voltage (IV) characteristics thus obtained for several values of  $U$, are shown in Fig \ref{fig:current}(a). In addition to the suppression of the current, we can see the enhanced nonlinear behavior. In order to observe the nonlinearity in more detail, we show the conductance, $I/V$, near the zero bias voltage in Fig \ref{fig:current}(b). In this figure, the conductance curve for $U/\Gamma_{N}=0$ is almost flat near the zero bias voltage, implying that the linear response theory can be safely applied in this finite voltage region. For $U/\Gamma_{N}=5$, the conductance is suppressed, yet keeps the flat structure. With further increase in $U$, however, the conductance shows a convex curve; the linear response regime is restricted to the very tiny voltage region, i.e., $|eV|/\Delta \lesssim 0.01$ for $U/\Gamma_{N}=20$. Therefore, theoretical studies only on the zero bias conductance are not enough to understand the transport properties in the actual experiments in the strong Coulomb interaction regime.

In the case of $\epsilon_{d}/U=-0.5$, we have an alternative expression for the zero bias conductance at absolute zero in terms of the renormalized couplings as~\cite{CuevasJC01kei, TanakaY07nrg}
\begin{equation}
\frac{\mathrm{d}I}{\mathrm{d}V }\Big|_{V=0} = \frac{16e^2}{h} \frac{\left(\widetilde{\Gamma}_{S}/\widetilde{\Gamma}_{N}\right)^2}{\left(1+\left(\widetilde{\Gamma}_{S}/\widetilde{\Gamma}_{N}\right)^2\right)^{2}},
\label{eq:lincnd_rp}
\end{equation}
where $\widetilde{\Gamma}_{N}$ and $\widetilde{\Gamma}_{S}$ are the renormalized couplings defined by
\begin{eqnarray}
\widetilde{\Gamma}_{N}&=&z\Gamma_{N},\label{eq:rGn}\\
\widetilde{\Gamma}_{S}&=&z(\Gamma_{S}+[\bm{\Sigma}_{U}^{r}(0)]_{12}),\label{eq:rGs}\\
z&=&(1+\frac{\Gamma_{S}}{|\Delta|}-\frac{\mathrm{d}[\bm{\Sigma}_{U}^{r}(\omega)]_{11}}{\mathrm{d}\omega}\Big|_{\omega=0})^{-1}. \label{eq:renormalization_factor}
\end{eqnarray}
Here, $[\bm{\Sigma}_{U}^{r}]_{ij}$ denotes the $(i, j)$ component of the Nambu matrix of the retarded self-energy at the QD. It is worthwhile to note that $\widetilde{\Gamma}_{N}$ and $\widetilde{\Gamma}_{S}$ can be calculated only from the retarded (or advanced) self-energy in the equilibrium states and there is no need to calculate the lesser and greater self-energies.

\begin{figure}
\includegraphics[width=8cm,clip]{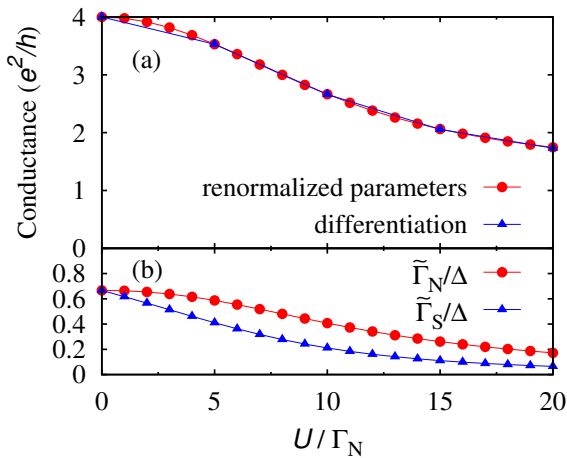}
\caption{\label{fig:lincnd_chk} (Color online) (a) Zero bias conductance as a function of  $U$ obtained from the two different methods for $\Gamma_{S}/\Gamma_{N}=1$, $\epsilon_{d}/U=-0.5$, $\Delta/\Gamma_{N}=0.5$, $k_{B}T/\Gamma_{N}=0$, which are respectively denoted by circles (scheme of renormalized parameters) and triangles (scheme of direct differentiation).
(b) Renormalized couplings $\widetilde{\Gamma}_{N}$ and $\widetilde{\Gamma}_{S}$. The parameters used are the same as in (a).}
\end{figure}
Figure \ref{fig:lincnd_chk}(a) shows the zero bias conductance obtained in the above-mentioned two different ways. The values of the conductance obtained from the differentiation (triangle) well coincide with those obtained from Eq.\eqref{eq:lincnd_rp} with the renormalized parameters (circle). This fact confirms the consistency of our MPT treatment at least around the zero bias voltage. The consistency is assured by the effective parameter $\lambda$ introduced for the current conservation in the MPT framework. As pointed out in the previous section, a simple second order perturbation may break the current conservation law; there are two different definitions of the current, $I_{N}$ and $I_{S}$. If the current conservation law is violated, at least one of the values of the zero bias conductance calculated from $I_{N}$ and $I_{S}$ is not consistent with the one obtained with the renormalized parameters. Therefore, the confirmation done here is important to obtain the sensible  results for the current at finite bias voltage.

In Fig. \ref{fig:lincnd_chk}(a), the zero bias conductance decreases with increasing $U$, which is due to the suppression of the Andreev reflection by the Coulomb interaction. The suppression of the Andreev reflection between the QD and the S-lead and also the single-electron tunneling between the QD and the N-lead are seen in the Coulomb interaction dependence of the renormalized couplings (Fig. \ref{fig:lincnd_chk}(b)). With increasing $U$, $\widetilde{\Gamma}_{S}$ decreases more rapidly than $\widetilde{\Gamma}_{N}$, indicating that the entire N-QD-S system is approximately decoupled into two parts in the low energy region: S-lead and QD-N systems. Therefore, the Kondo singlet state becomes dominant in the ground state of the QD for large $U$, leading to the suppression of the Andreev reflection.
\begin{figure}
\includegraphics[width=8cm,clip]{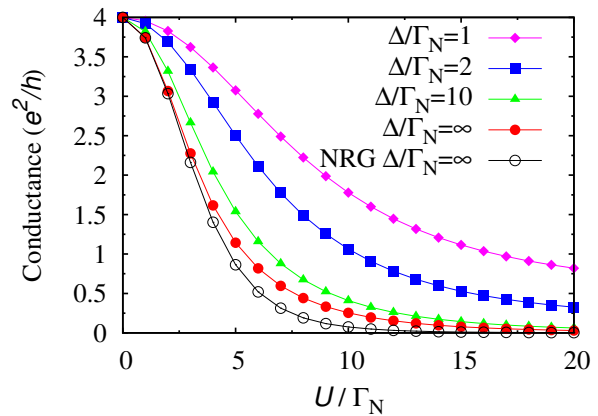}
\caption{\label{fig:lincnd_cmpNRG} (Color online) Zero bias conductance as a function of  $U$ for several values of $\Delta$. The closed and open symbols denote the results of MPT and NRG. The other parameters are  $\Gamma_{S}/\Gamma_{N}=1$, $\epsilon_{d}/U=-0.5$ and $k_{B}T/\Gamma_{N}=0$.}
\end{figure}

We next discuss the zero bias conductance in the large $\Delta$ region, in comparison with the results obtained with the numerical renormalization group (NRG) calculation~\cite{TanakaY07nrg}. Figure \ref{fig:lincnd_cmpNRG} shows the zero bias conductance as a function of $U$ for several values of  $\Delta$. We note that similar calculations have been done by Cuevas et al.~\cite{CuevasJC01kei} Let us first look at the case of infinitely large gap, where the closed and open circles denote the conductance obtained with MPT and NRG calculations. Although our approach is based on the perturbation expansion in $U$, the MPT results reproduce the NRG results in both weak and strong $U$ regions since in the MPT framework the effective parameters $\ave{\overline{n_{d}}}$ and $\ave{\overline{d_{\downarrow}d_{\uparrow}}}$ are self-consistently determined to reproduce the atomic limit correctly. Only in the intermediate region around $U/\Gamma_{N}=6$, we see some discrepancies between the two results (less than 0.3 $e^2/h$). 

Let us now discuss how the zero bias conductance depends on the superconducting gap $\Delta$ in Fig. \ref{fig:lincnd_cmpNRG}. With decreasing $\Delta$, the conductance for finite Coulomb interaction is enhanced. It should be noted that the renormalization factor $z$ is determined not only by $\Delta$ and $\Gamma_{S}$ but also by the retarded self-energy due to the Coulomb interaction, as seen in eq. \eqref{eq:renormalization_factor}; for finite $\Delta$ and $\Gamma_{S}$, $z$ is smaller than unity even in the noninteracting case. For small $\Delta$, the renormalization by the Coulomb interaction is weak, leading to the enhancement of the zero bias conductance, as compared with the case of $\Delta=\infty$.

\begin{figure}
\includegraphics[width=8cm,clip]{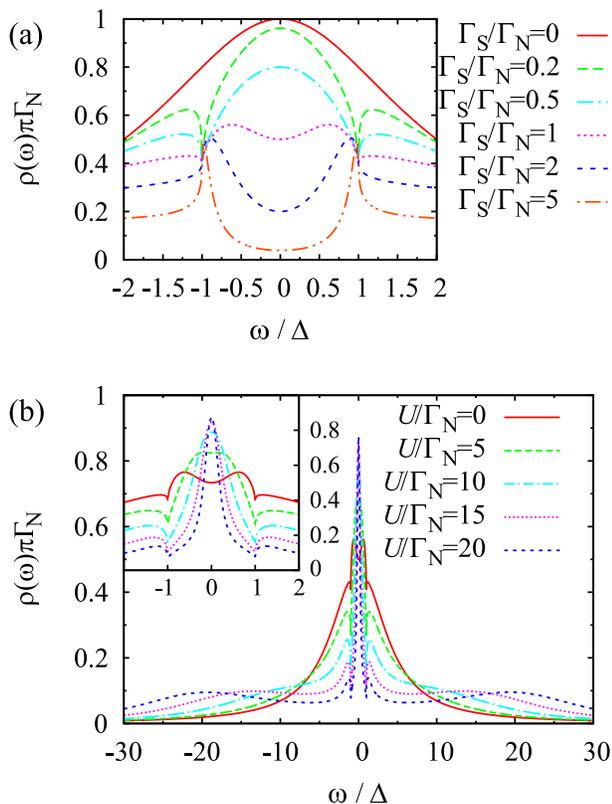}
\caption{\label{fig:dos_V0} (Color online) (a) LDOS for several values of $\Gamma_{S}$: $U=\epsilon_{d}=0$, $\Delta/\Gamma_{N}=0.5$ and $k_{B}T=0$. (b) LDOS for several values of $U$: $\Gamma_{S}/\Gamma_{N}=1$, $\epsilon_{d}/U=-0.5$, $\Delta/\Gamma_{N}=0.5$ and $k_{B}T=0$. The inset is the enlarged picture in the region around the Fermi energy.}
\end{figure}
We now look at the local density of states (LDOS) at the QD with and without the Coulomb interaction $U$.  The LDOS in the noninteracting case is shown in Fig. \ref{fig:dos_V0}(a). For $\Gamma_{S}=0$, there is a broad resonance around the Fermi energy due to the decoupling of the QD from the S-lead, which means that the QD is in the mixed valence regime. For small $\Gamma_{S}$, the weight of the LDOS is suppressed at $\omega=\pm\Delta$ since at the same energies, the divergence of DOS of S-lead occurs. With further increasing $\Gamma_{S}$, the LDOS develops a pseudo gap at the Fermi energy and a double-peak structure appears inside the gap owing to the superconducting proximity effect; the superconducting (SC) singlet state becomes dominant at the QD. Since the two resonances inside the gap are reduced to the Andreev bound states for $\Gamma_{N}/\Gamma_{S}=0$, we here refer to them as the Andreev resonances. The Andreev resonances are located at $\omega \simeq \pm \widetilde{\Gamma}_{S}$ with the same width $\widetilde{\Gamma}_{N}$. The change from a single resonance to the Andreev resonances clearly characterizes the crossover in the dominant couplings at the QD, which occurs around $\Gamma_{S}/\Gamma_{N}=1$.

In Fig. \ref{fig:dos_V0}(b), the LDOS at $\Gamma_{S}/\Gamma_{N}=1$ (crossover regime) is shown for several choices of $U$. With the increase in $U$, the Andreev resonances are merged into a single resonance, indicating that the superconducting correlations are reduced by the strong Coulomb interaction and the Kondo correlations are enhanced instead; the Kondo singlet state dominates the SC singlet state at the QD. The broad peaks corresponding to the charge excitations are also observed at $\omega \simeq U/2$ for large $U$. The $U$ dependence of the LDOS in Fig. \ref{fig:dos_V0}(b) is consistent with the preceding MPT calculations by Cuevas et al.~\cite{CuevasJC01kei}, though they did not address its relationship to the Kondo and SC singlet states.

\begin{figure}
\includegraphics[width=8cm,clip]{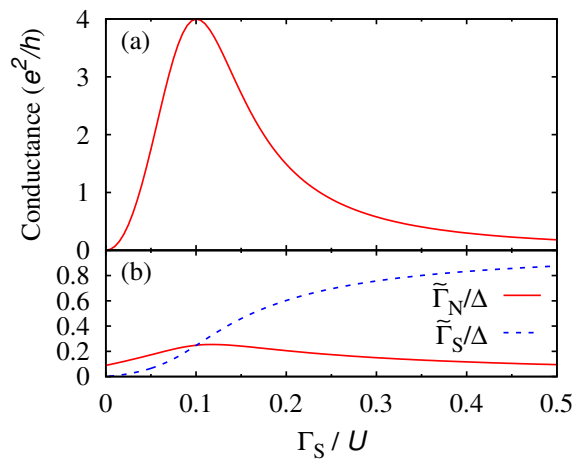}
\caption{\label{fig:lincnd_Gsvar} (Color online) (a) Zero bias conductance as a function of  $\Gamma_{S}$ for  $\Gamma_{N}/U=0.05$, $\epsilon_{d}/U=-0.5$, $\Delta/U=0.025$ and $k_{B}T/\Gamma_{N}=0$. (b) The renormalized parameters as a function of $\Gamma_{S}$.}
\end{figure}

In the particle-hole symmetric case ($\epsilon_{d}/U=-0.5$) with symmetric couplings $\Gamma_{S}/\Gamma_{N}=1$, the strong Coulomb interaction favors the Kondo singlet ground state, as discussed above. If we change the ratio $\Gamma_{S}/\Gamma_{N}$, however, the SC singlet state can be dominant in the ground state in the strong Coulomb interaction regime. Such examples are shown in Fig. \ref{fig:lincnd_Gsvar}, where the zero-bias conductance and the corresponding renormalized couplings are plotted as a function of $\Gamma_{S}$ for $\Gamma_{N}/U=0.05$.  For $\Gamma_{S}/U=\Gamma_{N}/U=0.05$, the N-lead is strongly coupled to the QD; $\widetilde{\Gamma}_{N}>\widetilde{\Gamma}_{S}$. As $\Gamma_{S}$ increases, $\widetilde{\Gamma}_{S}$ increases more rapidly than $\widetilde{\Gamma}_{N}$, although both of them are enhanced because the Coulomb interaction effects are suppressed by the superconducting proximity effects. We can indeed see that the crossover in the dominant couplings occurs around $\Gamma_{S}/U \simeq 0.1$. Further increase in $\Gamma_{S}$ leads to the enhancement of $\widetilde{\Gamma}_{S}$ and the suppression of $\widetilde{\Gamma}_{N}$, driving the system into the S-lead dominant coupling regime where the SC singlet is dominant at the QD.

Since the zero bias conductance has the maximum value for $\widetilde{\Gamma}_{S}/\widetilde{\Gamma}_{N}=1$ (see eq. \eqref{eq:lincnd_rp}), the conductance shows a peak structure around $\Gamma_{S}/U \simeq 0.1$ as shown in Fig. \ref{fig:lincnd_Gsvar}(a). Away from the crossover regime, the conductance decreases both in the N-lead and S-lead dominant coupling regimes. For any finite values of $\Gamma_{N}/U$ and $\Delta/U$, the crossover in the dominant couplings occurs at a finite $\Gamma_{S}/U$. We will see in the next section that the renormalized quantities $\widetilde{\Gamma}_{N}$ and $\widetilde{\Gamma}_{S}$ also characterize the differential conductance even at a finite bias voltage.

\subsection{Phase diagram}
\begin{figure}
\includegraphics[width=8cm,clip]{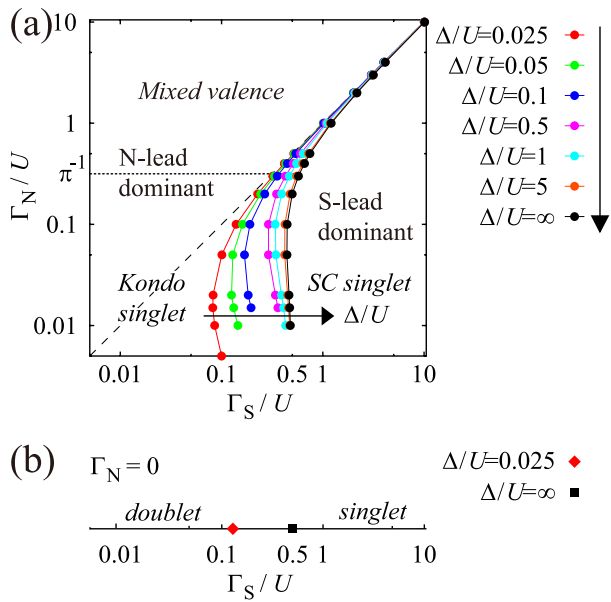}
\caption{\label{fig:phasediagram} (Color online) (a) Phase diagram for the particle-hole symmetric N-QD-S system in equilibrium conditions. The dominant state for finite $\Gamma_{N}$ and the ground state for $\Gamma_{N}=0$ at the QD are denoted in italic. (b) Phase diagram for the particle-hole symmetric QD-S system for $\Gamma_{N}=0$. The first order transition points of the ground state are denoted by the diamond ($\Delta/U=0.025$) and the square ($\Delta/U=\infty$).
}
\end{figure}
We summarize the results for the equilibrium N-QD-S system in the phase diagram specified in terms of the dominant couplings. Figure \ref{fig:phasediagram}(a) shows the phase diagram of the particle-hole symmetric N-QD-S system as functions of logarithms of $\Gamma_{N}/U$ and $\Gamma_{S}/U$. In this figure, the solid lines, which are determined by $\widetilde{\Gamma}_{N}=\widetilde{\Gamma}_{S}$, characterize the crossover behavior. In the left (right) region divided by the crossover line, the coupling between the QD and the N-lead (S-lead) is dominant, $\widetilde{\Gamma}_{N}>\widetilde{\Gamma}_{S}$ ($\widetilde{\Gamma}_{N}<\widetilde{\Gamma}_{S}$). 

First, let us look at the case of $\Delta=\infty$ in Fig. \ref{fig:phasediagram}(a). For large $\Gamma_{N}/U$, the crossover line approaches $\Gamma_{N}/\Gamma_{S}=1$ denoted by the dotted line because the effects of the Coulomb interaction become weak there and then $\widetilde{\Gamma}_{N}/\widetilde{\Gamma}_{S} \simeq \Gamma_{N}/\Gamma_{S}$ (see eqs. \eqref{eq:rGn} and \eqref{eq:rGs}). In the region of large $\Gamma_{N}/U$ but small $\Gamma_{S}/U$, the QD is strongly coupled only to the N-lead and the Coulomb interaction is week, so that the QD is in a mixed-valent singlet state. With increasing $\Gamma_{S}$, the proximity effects are enhanced and a pseudogap is formed in the LDOS at the QD, thus leading to the SC dominant state at the QD. On the other hand, in the small $\Gamma_{N}/U$ region ($\pi \Gamma_{N}/U < 1$), the crossover line considerably deviates from the noninteracting one ($\Gamma_{N}/\Gamma_{S}=1$) due to the renormalization effects by the Coulomb interaction. For $\pi \Gamma_{N}/U<1$, the mixed valence state is gradually changed into the Kondo singlet state, so that the crossover from the Kondo singlet state to the SC singlet state occurs as $\Gamma_{S}/U$ increases for small $\Gamma_{N}/U$.

The crossover line terminates at $\Gamma_{S}/U=0.5$ in the limit of $\Gamma_{N}/U \to 0$ as shown in Fig.~\ref{fig:phasediagram}(b), where the crossover is changed to a doublet-singlet transition. This is because the N-QD-S system is completely divided into two parts at $\Gamma_{N}=0$, the N-lead and the QD-S system. This doublet-singlet transition is easily seen in the case of $\Delta = \infty$, where the effective Hamiltonian of the QD-S system is simplified since the coupling between the Bogoliubov quasiparticles and the QD vanishes; It has a single level with the superconducting pairing potential characterized by the hybridization $\Gamma_{S}$ as already noted in eq. \eqref{eq::HamiDinfdot}. The resulting effective Hamiltonian can be diagonalized by the Bogoliubov transformation, leading to four eigenstates: two singly occupied states with spin 1/2, $\ket{\uparrow}$ and $\ket{\downarrow}$, and two states with total spin 0 consisting of a linear combination of the doubly occupied and empty states. In the particle-hole symmetric case, the spin-0 singlet states are given by
\begin{eqnarray}
\ket{S1}&=&\frac{1}{\sqrt{2}}(\ket{0}-\ket{\uparrow\downarrow}),\\
\ket{S2}&=&\frac{1}{\sqrt{2}}(\ket{0}+\ket{\uparrow\downarrow}).
\end{eqnarray}
Note that the singly occupied states are degenerate (zero energy),
 and $\ket{S1}$ and $\ket{S2}$ have the different energies, $E_{S1}=\frac{U}{2}-\Gamma_{S}$ and $E_{S2}=\frac{U}{2}+\Gamma_{S}$. Therefore, the candidates for the ground state are the magnetic doublet state, $\ket{\sigma}$, and the SC singlet state, $\ket{S1}$. Either of these two states can be the  ground state depending on the parameters, and a first order transition occurs at $\Gamma_{S}/U=0.5$, which is denoted by the black square on the $\Gamma_{N}=0$ line in Fig.~\ref{fig:phasediagram}(b); for $\Gamma_{S}/U<0.5$ ($\Gamma_{S}/U>0.5$), the ground state is the doublet state (singlet state).
If $\Gamma_{N}$ is small but has a finite value, the local moment of the doublet ground state is screened by the electrons in the N-lead and the Kondo singlet state becomes the ground state. Therefore, the characteristic behavior in the crossover line for small $\Gamma_{N}/U$ reflects a remnant of the doublet-singlet transition at $\Gamma_{N}=0$.

When the superconducting gap $\Delta$ becomes finite, the system is not so much simplified because of the existence of the coupling between the quasiparticles in the S-lead and the QD even for $\Gamma_{N}=0$. Therefore, the competition between the Kondo effect and the superconducting proximity effect becomes important. Even in this case, there is still a doublet-singlet transition, which is confirmed by several authors in the problem of a magnetic impurity in superconductors~\cite{SodaT67sei, ShibaH69stt, Muller-HartmannE70tom, MatsuuraT77teo, SatoriK92nrg, YoshiokaT00nrg, BauerJ07spo} and the $0-\pi$ transition of the QD-Josephson junctions~\cite{GlazmanLI89rjc, ClerkAA00lop, VecinoE03jct, ChoiM-S04kea, SianoF04jct, OguriA04qpt, TanakaY07kei, MengT09sdo}. 
The transition point shifts toward lower $\Gamma_{S}$ with decreasing $\Delta$. We denote the transition point for $\Delta/U=0.025$, which is obtained with the NRG calculation, by the diamond in Fig. \ref{fig:phasediagram}(b). 

Summarizing, the ground state of our system is always in the singlet phase for finite $\Gamma_{N}$, where three different-type singlet regions are smoothly connected to each other via crossover behaviors. Only for $\Gamma_{N}=0$, there exists a transition between the singlet and doublet states. 

\subsection{Andreev bound states}

Here some comments are in order on the nature of excited states. We start with the $\Gamma_{N}=0$ and $\Delta=\infty$ case. Since there are only four discrete eigenstates at the QD in this case as discussed above, the excited states are localized at the QD. For $\Gamma_{S}/U>0.5$, $\ket{\sigma}$ becomes the first excited state with the energy $\omega_{b}=|E_{S1}|$. In contrast, for $\Gamma_{S}/U<0.5$, $\ket{S1}$ becomes the excited state with the energy $\omega_{b}$. The other singlet state, $\ket{S2}$, is always the second excited state with the energy $\omega_{b2}=E_{S2}-\min(E_{S1},0)$ which is larger than $\omega_{b}$. The one particle excitation from the ground state to the excited states  localized at the QD is observed as sharp peaks in the LDOS. These sharp peaks correspond to the Andreev bound states. Therefore, there may be four Andreev bound states in the LDOS when the ground state is a magnetic doublet with energy $\pm \omega_{b}$ and $\pm \omega_{b2}$. On the other hand, when $\ket{S1}$ becomes the ground state, there are only two peaks with energy $\pm \omega_{b}$ because there is no one particle excitation from $\ket{S1}$ to $\ket{S2}$.

In the finite $\Delta$ case, there still exist the Andreev bound states inside the gap which correspond to $\ket{\sigma}$ or $\ket{S1}$ for any values of $\Delta$ though the energy of the bound states $\omega_{b}$ cannot be obtained easily. Moreover, the second excited state corresponding to $\ket{S2}$ may be outside of the gap and be absorbed into the continuum energy spectrum for small $\Delta$~\cite{MengT09sdo}. In that case, there are only two Andreev bound states at the QD even though the magnetic doublet state is the ground state.

In the recent experiments~\cite{DeaconRS10kat,DeaconRS10tso}, Deacon {\it et al.} have found that the Andreev bound states at the QD can be detected in the nonequilibrium transport measurements in an N-QD-S system. In the experiments, there are only two kinds of peaks which correspond to two kinds of the first excited bound states. This fact may be attributed to the small superconducting gap prepared in the experiments. This issue will be addressed in the next section focusing on the nonequilibrium transport.

\section{ Nonequilibrium electron transport}

In this section, we study the nonequilibrium electron transport in the N-QD-S system with a special focus on the influence of the Kondo effect and the Andreev scattering on the nonlinear transport. In particular, we clarify the origin of the characteristic structures in the conductance profile in comparison with the recent experiments.

\begin{figure}
\includegraphics[width=9cm,clip]{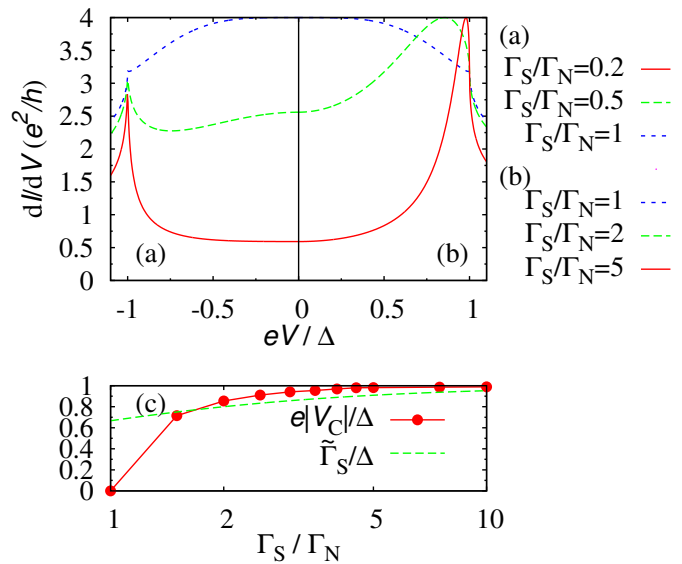}
\caption{\label{fig:cnd_GsV_U0} (Color online) (a), (b) Bias voltage dependence of the differential conductance for several values of $\Gamma_{S}$, $U/\Gamma_{N}=0$, $\epsilon_{d}=0$, $\Delta/\Gamma_{N}=0.5$ and $k_{B}T/\Gamma_{N}=0$. 
Here, we only show the results either for positive or negative $V$ since $\mathrm{d}I/\mathrm{d}V$ is symmetric with respect to $V=0$.
 (c) Peak position of $dI/dV$, $V_{C}$,  
in comparison with $\widetilde{\Gamma}_{S}$.}
\end{figure}

Before elucidating the Coulomb interaction effects on the nonequilibrium electron transport, it is instructive to discuss the differential conductance, $dI/dV$, in the noninteracting case. Here, we set $\epsilon_{d}=0$, where $dI/dV$ has a symmetric profile with respect to the $V=0$ axis. Figure \ref{fig:cnd_GsV_U0} shows the differential conductance for several values of $\Gamma_{S}/\Gamma_{N}$ as a function of the positive or negative bias voltage $V$.

In the noninteracting case, the ratio of the bare couplings directly determines the nature of the system. We can see the $dI/dV$ profiles characteristic in the N-lead and S-lead dominant coupling regimes in Fig. \ref{fig:cnd_GsV_U0}(a) and (b), respectively. In the symmetric coupling case, $\Gamma_{S}/\Gamma_{N}=1$, the differential conductance has a maximum value $4e^2/h$ at $V=0$. The zero bias conductance is suppressed both for $\Gamma_{S}/\Gamma_{N}< 1$ and $\Gamma_{S}/\Gamma_{N}> 1$. In the nonlinear regime, however, the differential conductance in the two regimes has different characteristics.

In the N-lead dominant coupling regime as seen in Fig. \ref{fig:cnd_GsV_U0}(a), the differential conductance is suppressed in whole subgap voltages with decreasing $\Gamma_{S}$ because of the suppression of the Andreev reflection. In the $\Gamma_{S} \to 0$ limit, the sharp peaks appear at the gap edges and the profile of $dI/dV$ becomes similar to the density of states in the S-lead, indicating that the system in the limit approaches the one with a NS tunnel junction~\cite{KhlusVA93rtt}. On the other hand, in the S-lead dominant coupling regime shown in Fig \ref{fig:cnd_GsV_U0}(b), the peak of zero bias conductance is split into two, which then move toward the opposite gap edges with keeping the unitary-limit value of $4e^2/h$ when $\Gamma_{S}$ increases. In this case, the SC-singlet is dominant at the QD in the equilibrium state, and the Andreev resonances, which originate from the exited doublet state for $\Gamma_{N}=0$, emerge in the LDOS at the QD as shown in Fig. \ref{fig:dos_V0}(a). The positions of the resonances are approximately given by $\pm \widetilde{\Gamma}_{S}$. We compare the voltage $V_{C}$ that gives a peak in $dI/dV$ with $\widetilde{\Gamma}_{S}$ in Fig. \ref{fig:cnd_GsV_U0}(c). It is seen that $eV_{C}$ moves along the curve of $\widetilde{\Gamma}_{S}$ for large $\Gamma_{S}/\Gamma_{N}$, which confirms that the subgap peak in $dI/dV$ results from the enhancement of the transport through the Andreev resonances.

\subsection{Nonequilibrium transport for particle-hole symmetric case: $\epsilon_{d}/U=-0.5$ }

We now investigate the Coulomb interaction effects on the nonequilibrium differential conductance at a finite bias voltage. Since there are many relevant parameters in the system, we will divide our discussions into two cases. We first treat the simple case with a condition of particle-hole symmetry, $\epsilon_{d}/U=-0.5$. More generic cases with arbitrary conditions for $\epsilon_{d}$ and $U$ will be discussed separately in the next subsection in comparison with the experiments.

\subsubsection{Coulomb interaction effects for $\Gamma_{S}/\Gamma_{N}=1$}

\begin{figure}
\includegraphics[width=8cm,clip]{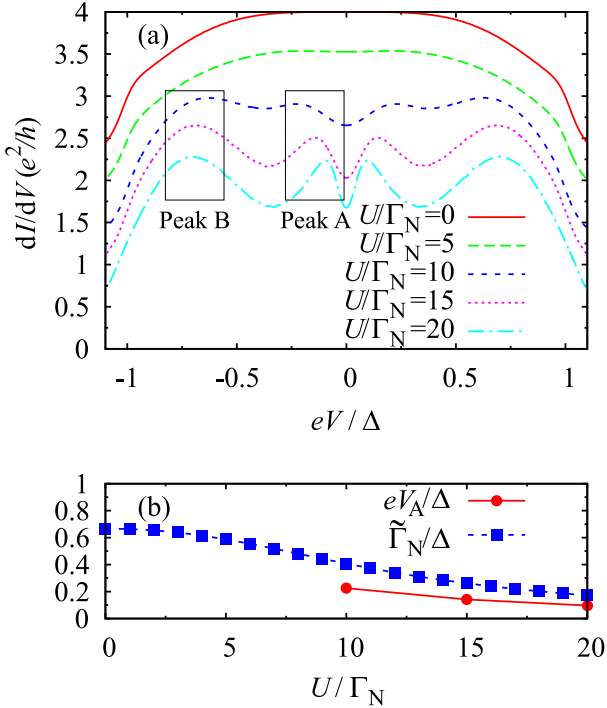}
\caption{\label{fig:cnd_UV} (Color online) (a) Bias voltage dependence of the differential conductance for several values of $U$: $\Gamma_{S}/\Gamma_{N}=1$, $\epsilon_{d}/U=-0.5$, $\Delta/\Gamma_{N}=0.5$ and $k_{B}T/\Gamma_{N}=0.01$. (b) Plots of the peak position, $V_{A}$, in the $dI/dV$ curve near the zero bias voltage and the renormalized coupling, $\widetilde{\Gamma}_{N}$, which is calculated for $V=T=0$.}
\end{figure}

Let us start with a system with the symmetric couplings for tunneling, $\Gamma_{S}/\Gamma_{N}=1$, which may help us to imagine what is essential in the nonequilibrium transport in the interacting QD.  Figure \ref{fig:cnd_UV}(a) shows the differential conductance as a function of the bias voltage for several values of the Coulomb interaction, $U$. According to the analysis in the previous section (see Fig. \ref{fig:phasediagram}), with increasing $U$, the system enters the N-lead dominant coupling regime where the Kondo singlet state becomes dominant. In this Kondo regime, several peaks appear at subgap voltages. We refer to the two sharp peaks near the zero bias voltages as Peak A and the two broad peaks at higher voltages as Peak B in Fig. \ref{fig:cnd_UV}(a). Although the heights of the peaks are suppressed, both of Peak A and Peak B become prominent for large $U$. Note that the $U$ dependence of the position of Peak A is different from that of Peak B; Peak A approaches the zero bias voltage with increasing $U$, whereas Peak B slightly shifts toward the gap edge. This fact implies that these two kinds of subgap peaks in $dI/dV$ have different origins. We will show below that Peak A originates from the interplay between the Kondo effect and the Andreev reflection at a finite bias, while Peak B comes from the Andreev bound states at the QD. 
\begin{figure}
\includegraphics[width=8cm,clip]{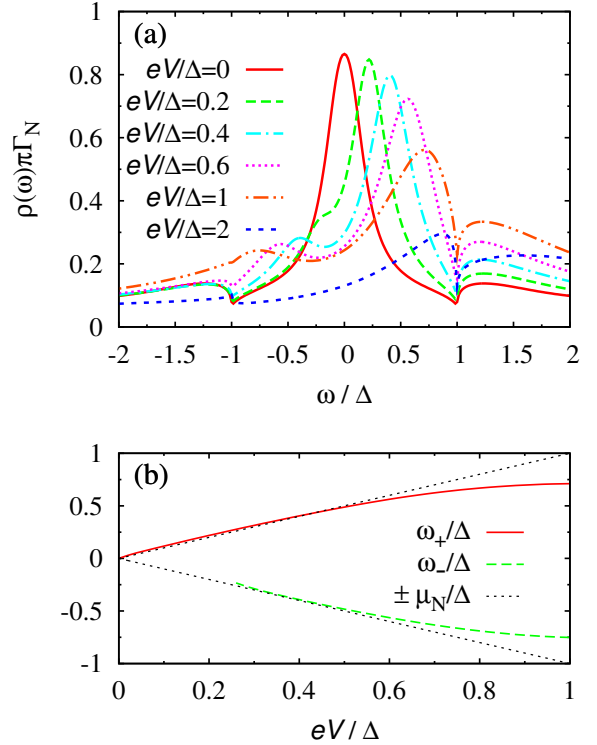}
\caption{\label{fig:dos_Gs1} (Color online) (a) Local density of states at the QD for several values of $V$: $\Gamma_{S}/\Gamma_{N}=1$, $U/\Gamma_{N}=20$, $\epsilon_{d}/U=-0.5$, $\Delta/\Gamma_{N}=0.5$ and $k_{B}T/\Gamma_{N}=0.01$. (b) Peak position of subgap resonances in (a) as a function of $V$.
}
\end{figure}

Let us focus on Peak A. Figure \ref{fig:cnd_UV}(b) shows the comparison of the position of Peak A, denoted as $V_{A}$, and the renormalized N-lead coupling $\widetilde{\Gamma}_{N}$ defined at $V=T=0$. It is seen that the value of $eV_{A}$ approaches $\widetilde{\Gamma}_{N}$ when the system enters the Kondo regime with increasing $U$. Since $\widetilde{\Gamma}_{N}$ is the characteristic energy scale of the Kondo effect, which approximately gives the width of the Kondo resonance, Peak A is related to the Kondo effect. The emergence of the Kondo effect is seen in the bias voltage dependence of the LDOS at the QD shown in Fig. \ref{fig:dos_Gs1}. Although the LDOS for $U=0$ is not changed by the bias voltage, it is affected via the self-energy for finite $U$. In particular, the LDOS in the Kondo regime substantially changes its form under a finite bias voltage. Figure \ref{fig:dos_Gs1}(a) shows the LDOS at the QD for $U/\Gamma_{N}=20$ and $\Gamma_{S}/\Gamma_{N}=1$. For $V=0$, there is a sharp Kondo resonance at the Fermi energy. With increasing $V$, the position of the Kondo resonance follows the chemical potential of the N-lead, $\mu_{N}=eV$, suggesting that the Kondo screening of the local moment is mainly caused by the normal lead.

A noticeable change in the LDOS at finite bias voltage (Fig. \ref{fig:dos_Gs1}(a)) is the appearance of the additional resonance which is located at the counter position of the ordinary Kondo resonance; the ordinary Kondo resonance has a shoulder structure for $eV/\Delta=0.2$, which is changed into an additional resonance for $eV/\Delta=0.4$. This consideration naturally suggests that the additional resonance is caused by the Andreev reflection through the ordinary Kondo resonance (referred to as Kondo-assisted Andreev reflection); an electron which comes from the N-lead Fermi surface reaches the S-lead via the ordinary Kondo resonance, and then it is converted as a hole via the Andreev reflection process. Since the electron has finite energy measured from the S-lead Fermi surface, $eV$, the reflected hole also has the same energy. This interpretation clarifies why the position of the additional resonance is located at the counter position of the Kondo resonance.
Note that the additional resonance discussed here was previously realized by Sun {\it et al.}~\cite{SunQ-f01ekr}, but was not discussed in detail, in particular, about its physical relevance to the transport properties. We will address this issue with the use of the renormalized couplings, and demonstrate that it indeed provides a source of the marked change in nonequilibrium transport properties.

The positions of the Kondo and additional resonances, $\omega_{\mathrm{+}}$ and $\omega_{\mathrm{-}}$ are shown in Fig. \ref{fig:dos_Gs1}(b) as a function of the bias voltage. For small $V$, $\omega_{\mathrm{+}}$ and $\omega_{\mathrm{-}}$ follow the dotted lines which denote the position of the chemical potential of N-lead and its counter position, $\pm\mu_{N}$. Hence the crossover of the LDOS from the single peak to the double peaks occurs at $eV \simeq \widetilde{\Gamma}_{N}$ where the distance between the two peaks is approximately given by their width. By comparing the results of the LDOS with $dI/dV$, we find that the crossover voltage in the LDOS approximately corresponds to the one giving Peak A in the differential conductance. Summarizing all these results, we conclude that Peak A in $dI/dV$ originates from the Kondo-assisted Andreev reflection at a finite bias voltage.

\begin{figure}
\includegraphics[width=8cm,clip]{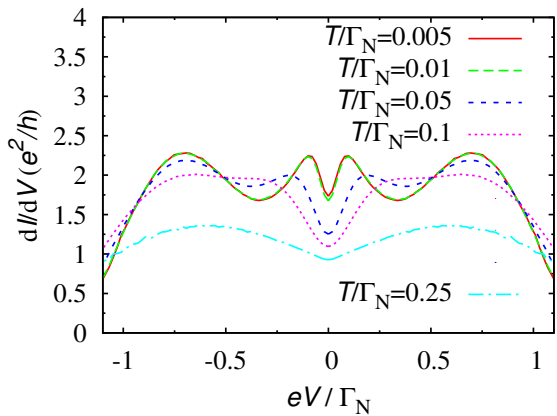}
\caption{\label{fig:cnd_TdepGs1} (Color online) Temperature dependence of the differential conductance for $\Gamma_{S}/\Gamma_{N}=1$, $U/\Gamma_{N}=20$, $\epsilon_{d}/U=-0.5$ and $\Delta/\Gamma_{N}=0.5$.
}
\end{figure}

In order to further confirm our interpretation for Peak A in $dI/dV$, we calculate the temperature dependence of $dI/dV$ as shown in Fig. \ref{fig:cnd_TdepGs1}. With increasing temperature, two peaks near  zero bias voltage, which are classified as Peak A, are smeared and absorbed into the broad peaks of Peak B. The temperature dependence of Peak A supports that it is due to the Kondo-assisted Andreev reflection. The characteristic temperature around which Peak A is smeared  coincides approximately with $\widetilde{\Gamma}_{N}/\Gamma_{N} \simeq 0.086$. We note that the equilibrium quantity of $\widetilde{\Gamma}_{N}$ characterizes both the position and the $T$-dependence of Peak A in the nonlinear differential conductance.

In contrast, we can see that Peak B in $dI/dV$ is not directly related to the Kondo effect according to its $T$-dependence in Fig. \ref{fig:cnd_TdepGs1}; although the two peaks labeled as Peak B decrease their heights with increasing $T$, the broad peak structure still exists even at $T/\Gamma_{N}=0.25$ (higher than the Kondo temperature). We indeed find that Peak B is solely controlled by the Andreev reflection, but not by the Kondo effect. Since the origin and the physical implications of Peak B are naturally seen in the asymmetric limit of $\Gamma_{S}/\Gamma_{N} \gg 1 $, we will discuss the physical properties systematically in asymmetric couplings below.

\subsubsection{Coulomb interaction effects for $\Gamma_{S}/\Gamma_{N} \neq 1$}

\begin{figure}
\includegraphics[width=8cm,clip]{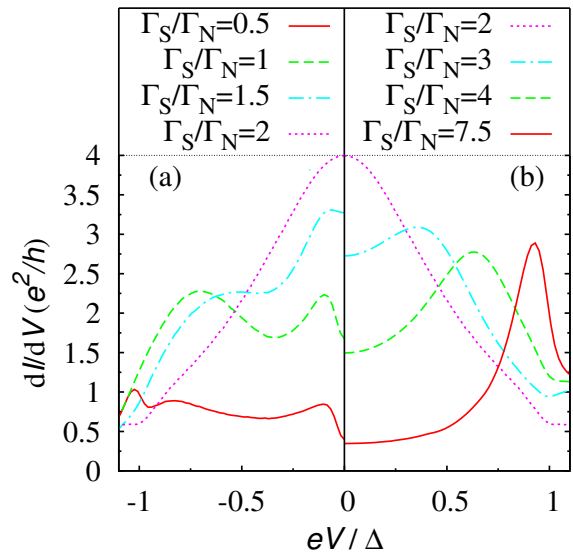}
\caption{\label{fig:cnd_GsV} (Color online) Differential conductance as a function of bias voltage for several values of $\Gamma_{S}$: $U/\Gamma_{N}=20$, $\epsilon_{d}/U=-0.5$, $\Delta/\Gamma_{N}=0.5$ and $k_{B}T/\Gamma_{N}=0.01$.}
\end{figure}

Here, we address how the asymmetry of couplings ($\Gamma_{S}/\Gamma_{N} \neq 1$) alters the nonequilibrium transport properties. We start with the differential conductance $dI/dV$ shown in Fig. \ref{fig:cnd_GsV} for several values of $\Gamma_{S}$  in the case of strong interaction $U/\Gamma_{N}=20$. Here, we only show the results either in positive or negative  $V$ since $\mathrm{d}I/\mathrm{d}V$ is symmetric with respect to $V=0$ for $\epsilon_{d}/U=-0.5$.  The crossover between the different regimes occurs around $\Gamma_{S}/\Gamma_{N} \simeq 2$ where $dI/dV$ has its maximum value $4e^2/h$ at zero bias voltage. The profiles of $dI/dV$ in the N-lead and S-lead dominant coupling regimes are shown in Fig. \ref{fig:cnd_GsV}(a) and (b), respectively.

In Fig. \ref{fig:cnd_GsV}(b), the $\mathrm{d}I/\mathrm{d}V$ curves show properties analogous to those in the noninteracting S-lead dominant coupling case shown in Fig. \ref{fig:cnd_GsV_U0}(b); the peak of $\mathrm{d}I/\mathrm{d}V$ moves toward the gap edge with increasing $\Gamma_{S}$. The peak values, however, depend on the Coulomb interaction and become smaller than $4e^2/h$ for large $U$. The suppression of the peak is attributed to the inelastic scattering owing to the Coulomb interaction. However, in the extreme limit of $\Gamma_{S} \gg \Gamma_{N}$, $U$, the superconducting correlation dominates the QD and thus the Coulomb interaction effects are reduced, leading to the suppression of the inelastic scattering. For $\Gamma_{S}/\Gamma_{N}=7.5$, therefore, the peak value increases again. For later discussions, we refer to these peaks as Peak C. On the other hand, 
the results of $\mathrm{d}I/\mathrm{d}V$ in the N-lead dominant coupling regime(Fig. \ref{fig:cnd_GsV}(a)) are a little bit complicated; the peak at zero bias voltage splits with decreasing $\Gamma_{S}$, forming a double-peak structure both in the positive and negative half of the subgap voltage regions. The origin of these two peaks is the same as discussed above, so that we denote them as Peak A and Peak B. With decreasing $\Gamma_{S}$, the position of Peak A hardly changes, while that of Peak B shifts toward the gap edge. With further decreasing $\Gamma_{S}$, both of Peak A and Peak B reduce their heights. Therefore, for $\Gamma_{S} \to 0$, $dI/dV$ in the gap is completely suppressed as is the case for $U=0$.

\begin{figure}
\includegraphics[width=8cm,clip]{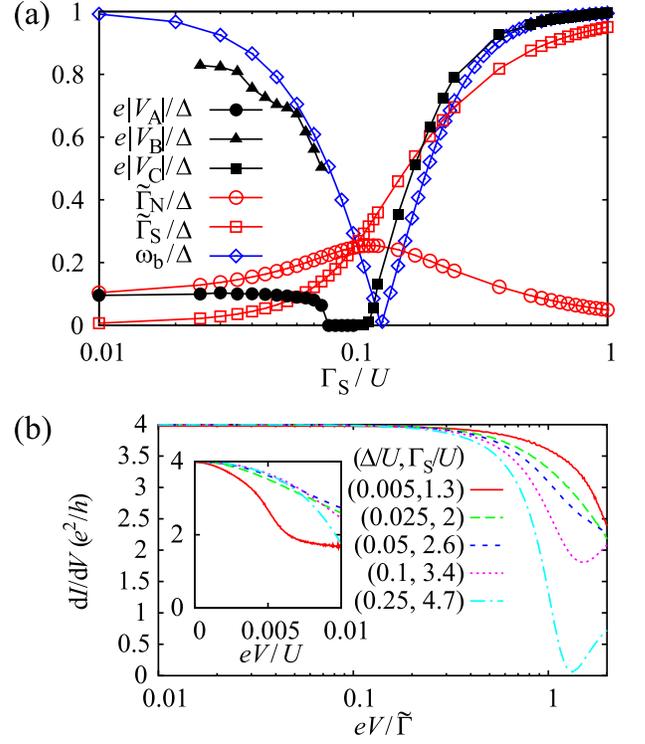}
\caption{\label{fig:peak_compare} (Color online) (a) Semilog plot of the peak positions of $\mathrm{d}I/\mathrm{d}V$ in the N-QD-S system, $V_{A}$ and $V_{B}$, as a function of $\Gamma_{S}/U$. The other parameters are $\Gamma_{N}/U=0.05$, $\epsilon_{d}/U=-0.5$, $\Delta/U=0.025$ and $k_{B}T/\Gamma_{N}=0.01$, which are the same as in Fig. \ref{fig:cnd_GsV}. For comparison, the renormalized couplings in the N-QD-S system with $V=T=0$, $\widetilde{\Gamma}_{N}$ and $\widetilde{\Gamma}_{S}$, and the energy of the Andreev bound states in the QD-S system with $T=0$, $\omega_{b}$, are also plotted. (b) Semilog plot of $dI/dV$ at the crossover point against $V/\widetilde{\Gamma}$ for several choices of $\Delta$ and $\Gamma_{S}$. The other parameters are the same as in (a). The value of $\widetilde{\Gamma}$ is $0.0551$, $0.124$, $0.157$, $0.187$ and $0.213$ for each curve from top to bottom. In the low voltage region with $eV<\widetilde{\Gamma}$, $dI/dV$ fall into a single curve. Inset shows the same data of $dI/dV$ as a function of $eV/U$.}
\end{figure}

In order to elucidate the origin of the peak formation, we plot the $\Gamma_{S}$ dependence of the positions of Peak A, B and C,  which are labeled as $V_{A}$, $V_{B}$ and $V_{C}$, in Fig. \ref{fig:peak_compare}(a). Here, we take $U$ as the energy unit. For comparison, the renormalized couplings, $\widetilde{\Gamma}_{N}$ and $\widetilde{\Gamma}_{S}$, are also plotted. 
For $\Gamma_{S}/U \lesssim (\gtrsim) 0.1$, $\widetilde{\Gamma}_{N}$ is larger (smaller) than $\widetilde{\Gamma}_{S}$, namely, the system is in the N-lead (S-lead) dominant coupling regime. It is seen that $V_{A}$ and $V_{C}$ approach the values of $\widetilde{\Gamma}_{N}$ and $\widetilde{\Gamma}_{S}$ in the limit of $\Gamma_{S} \rightarrow 0$ and $\infty$, respectively. 
 The crossover between these two limits appears around $\Gamma_{S}/U \simeq 0.1$, where the peak is located at $V=0$ with the unitary limit value, $4e^2/h$. At the crossover point, the width of the zero bias peak is simply scaled by $\widetilde{\Gamma}=\widetilde{\Gamma}_{N}=\widetilde{\Gamma}_{S}$ as shown in Fig. \ref{fig:peak_compare}(b). It is seen that the differential conductance calculated for several choices of $\Delta$ and $\Gamma_{S}$  quickly decreases around $|eV| = \widetilde{\Gamma}$.

The above properties in the conductance are clearly understood in terms of the Andreev bound states. The open diamonds in Fig. \ref{fig:peak_compare}(a) denote the energy $\omega_{b}$ of the Andreev bound states at the QD for $\Gamma_{N}=T=0$, which is obtained with the NRG calculations~\cite{SatoriK92nrg, YoshiokaT00nrg, BauerJ07spo}. As mentioned in the previous section, the system shows a transition between the magnetic doublet and SC singlet states for $\Gamma_{N}=T=0$. The transition point is evaluated as $\Gamma_{S}^{TP}/U \simeq 0.129$ from the condition $\omega_{b}=0$. For $\Gamma_{S} \le(\ge) \Gamma_{S}^{TP}$, the doublet (SC singlet) becomes the ground state and the Andreev bound states originating from the SC singlet (doublet) appear in the LDOS at the QD (see also the discussion in Fig. \ref{fig:phasediagram}). For finite $\Gamma_{N}$, the local moment of the doublet ground state is screened by the electrons in the N-lead. Therefore, the transition change into the crossover between the Kondo singlet and the SC singlet. It is clearly seen in Fig. \ref {fig:peak_compare}(a) that $V_{B}$ is indeed related to the Andreev bound states since $V_B$ approximately coincides with the energy of the Andreev bound states. This is also the case for Peak C, and the difference between them comes from whether the ground state is the Kondo singlet (Peak B) or SC singlet (Peak C). We therefore reveals the origin of Peak B and C; when the energy corresponding to the Andreev bound states is externally supplied by the applied bias voltage, the weight of the excited state at the QD is increased, resulting in the enhancement of the Andreev reflection. Although the nontrivial correspondence between the peak position of $dI/dV$ and the energy of the Andreev bound states has already been discovered by Deacon {\it et al}\cite{DeaconRS10tso, DeaconRS10kat} via the experimental studies, to our knowledge, this is the first numerical calculation which systematically clarifies the correspondence in both coupling regimes by taking into account the Kondo effect. 

\begin{figure}
\centering
\includegraphics[width=8cm,clip]{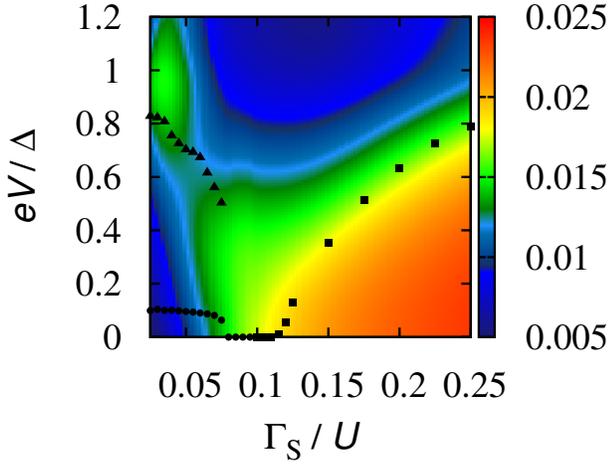}
\caption{\label{fig:Psi_GsV} (Color online) False color-scale representation of $|\langle d_{\downarrow}d_{\uparrow}\rangle |$ as a function of $V$ and $\Gamma_{S}$ for $\Gamma_{N}/U=0.05$, $\epsilon_{d}/U=-0.5$, $\Delta/\Gamma_{N}=0.5$ and $k_{B}T/\Gamma_{N}=0.01$. The filled circles, triangles and squares indicate the $\Gamma_{S}$ dependence of $eV_{A}/\Delta$, $V_{B}/\Delta$ and $V_{C}/\Delta$.}
\end{figure}

The difference between $V_B$ and $V_C$ can be more clearly seen in the superconducting pairing correlation at the QD, $|\langle d_{\downarrow}d_{\uparrow}\rangle |$.  Figure \ref{fig:Psi_GsV} shows the false color-scale representation of $|\langle d_{\downarrow}d_{\uparrow}\rangle |$ as a function of $\Gamma_{S}$ and $V$. The filled circles, triangles and squares on the representation indicate the $\Gamma_{S}$ dependence of $eV_{A}/\Delta$, $eV_{B}/\Delta$ and $eV_{C}/\Delta$ shown in Fig. \ref{fig:peak_compare}. In the N-lead dominant coupling regime, e.g., in the case of $\Gamma_{S}/U=0.05$, the superconducting correlation at the QD is weak at $V=0$ since the Kondo singlet is dominant. With increasing $V$, $|\langle d_{\downarrow}d_{\uparrow}\rangle |$ shows the peak at the bias voltage where Peak B is located. This result clearly indicates the enhancement of the weight of the SC singlet state at the finite bias voltage. On the other hand, in the S-lead dominant coupling regime, e.g., in the $\Gamma_{S}/U=0.3$ case, it is seen that $|\langle d_{\downarrow}d_{\uparrow}\rangle |$ is relatively large at $V=0$, and monotonically decreases with increasing $V$. In particular, a rapid decrease of $|\langle d_{\downarrow}d_{\uparrow}\rangle |$ occurs around $V=V_{C}$, implying that the weight of the magnetic doublet state increases in the system instead of the SC singlet state at $V=V_{C}$. All these features are consistent with the above interpretation of the peaks in the conductance.

\begin{figure}
\includegraphics[width=8cm,clip]{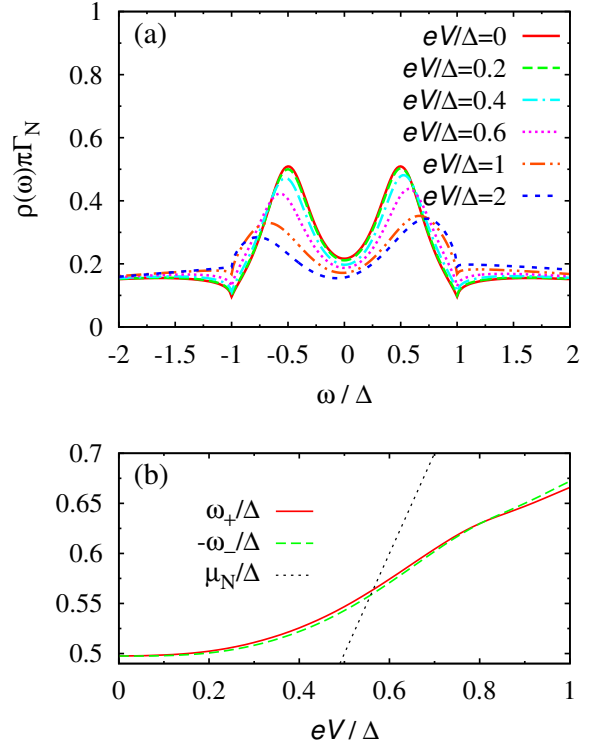}
\caption{\label{fig:dos_Gs3} (Color online) (a) Local density of states at the QD for several values of $V$ for $\Gamma_{S}/U=0.15$, $\Gamma_{N}/U=0.05$, $\epsilon_{d}/U=-0.5$, $\Delta/\Gamma_{N}=0.025$ and $k_{B}T/\Gamma_{N}=0.01$.}
\end{figure}

It is instructive to consider the LDOS for finite $V$ to see the nature of the bound states. Since the $V$ dependence of the LDOS in the N-lead dominant coupling regime has already been discussed in Fig. \ref{fig:dos_Gs1}, here we focus on the S-lead dominant coupling regime. Figure \ref{fig:dos_Gs3}(a) shows the LDOS at the QD for $\Gamma_{S}/U=0.15$. In this S-lead dominant coupling case, there are the Andreev resonances corresponding to the excited doublet for $\Gamma_{N}=0$. The double peaks of the LDOS in the equilibrium state are located at $\omega/\Delta \simeq \pm \widetilde{\Gamma}_{S} \simeq \pm 0.5$. For $eV/\Delta=0.2$, the resonances show little change as if the Coulomb interaction is absent in the system. For $eV/\Delta > 0.4$, however, the bias voltage increases the distance between the two resonances and smears them. In order to see the $V$ dependence of the resonances in detail, we plot the peak position of the resonances as a function of $V$ in Fig. \ref{fig:dos_Gs3}(b). Here, $\omega_{+}$ and $\omega_{-}$ denote the positions of the peaks for positive and negative $\omega$, respectively. With increasing $V$, they move toward the opposite gap edges. In particular, the peaks become sensitive to the change of the bias voltage around $eV/\Delta \simeq 0.3$, which approximately corresponds to the value of $eV_{C}/\Delta$. This feature indicates that the peaks tend to follow the chemical potential of N-lead and its counter position, $\mu_{N}$ and $-\mu_{N}$, which can be regarded as a kind of pinning effect of the Andreev resonances. Note that the profiles of $\omega_{+}$ and $-\omega_{-}$  approximately coincide with each other, implying that the resonances keep their symmetric structure with respect to the Fermi level of S-lead. Besides, in the N-lead dominant coupling regime, the pinning of the resonance becomes more prominent as discussed in Fig. \ref{fig:dos_Gs1}. The origin of the pinning is attributed to the Kondo effect, and is understood as follows: in the S-lead dominant coupling case, the electron correlation is practically negligible in the low energy and low voltage region, except for the renormalization effects, since the SC singlet is dominant at the QD. With increasing $V$, however, the weight of the magnetic doublet state increases in the system near $V \simeq V_{C}$. Then the resulting doublet state is screened by the electrons in the N-lead owing to the Kondo effect, leading to the pinning of the resonances. 

\begin{figure}
\includegraphics[width=8cm,clip]{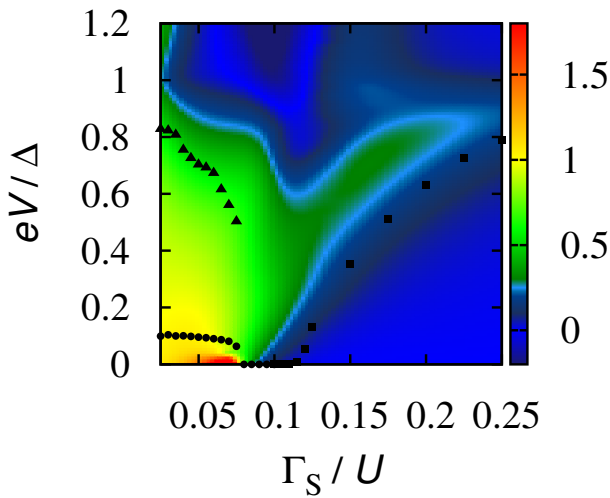}
\caption{\label{fig:dOdV_GsV} (Color online) False color-scale representation of $\mathrm{d}\omega_{+}/\mathrm{d}\mu_{N}$ as functions of $V$ and $\Gamma_{S}$ for $\Gamma_{N}/U=0.05$, $\epsilon_{d}/U=-0.5$, $\Delta/U=0.025$ and $k_{B}T/\Gamma_{N}=0.01$. The filled circles, triangles and squares indicate $eV_{A}/\Delta$, $V_{B}/\Delta$ and $V_{C}/\Delta$.}
\end{figure}

In order to further investigate the pinning of the resonances in the LDOS, we show the false color-scale representation of $\mathrm{d}\omega_{+}/\mathrm{d}\mu_{N}$ in Fig \ref{fig:dOdV_GsV}.  Since $\mathrm{d}\omega_{+}/\mathrm{d}\mu_{N}$ becomes large if the resonance of the LDOS in the positive $\omega$ region follows the chemical potential of the N-lead, its value gives an estimate of how strong the Kondo correlation is. In the N-lead dominant coupling regime, $\Gamma_{S}/U \lesssim 0.1$, the Kondo pinning effect is suppressed with increasing $V$ since the bias voltage destroys the Kondo singlet state. In particular, $\mathrm{d}\omega_{+}/\mathrm{d}\mu_{N}$ is rapidly decreases when the value of the bias voltage approaches $V_{B}$ because the weight of the SC singlet is increased at the voltage.  In contrast, in the S-lead dominant coupling regime, $\mathrm{d}\omega_{+}/\mathrm{d}\mu_{N}$ increases with increasing $V$ from zero and takes a peak at a finite bias voltage where Peak C appears in $dI/dV$. Therefore, it is intuitively understood that the Kondo correlation is enhanced by the bias voltage around $V=V_{C}$. With further increase in $V$, the Kondo correlation is weakened again by the applied bias voltage, which in turn leads to the suppression of $\mathrm{d}\omega_{+}/\mathrm{d}\mu_{N}$. The different features in $\mathrm{d}\omega_{+}/\mathrm{d}\mu_{N}$ at $V=V_{B}$ and $V=V_{C}$ reflect the difference in the origin of the Andreev bound states. Therefore, all the results of $dI/dV$, $|\langle d_{\downarrow}d_{\uparrow}\rangle |$ and $\mathrm{d}\omega_{+}/\mathrm{d}\mu_{N}$ are consistent with the scenario that the weight of the excited states is enhanced when the strength of the bias voltage  coincides with the energy of the excited states.

\subsubsection{Comparison with mean-field results}

\begin{figure}
\includegraphics[width=8cm,clip]{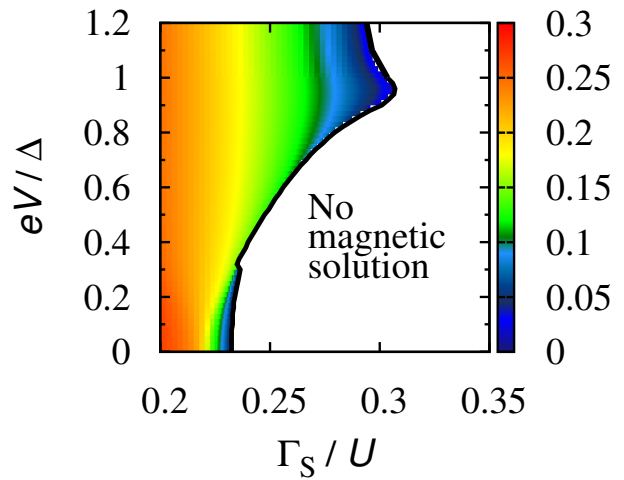}
\caption{\label{fig:mag} (Color online) False color-scale representation of the magnetization at the QD as functions of $V$ and $\Gamma_{S}$ for $\Gamma_{N}/U=0.05$, $\epsilon_{d}/U=-0.5$, $\Delta/U=0.025$ and $k_{B}T/\Gamma_{N}=0.01$.}
\end{figure}

In order to clarify the role of electron correlations at a finite bias voltage, it is instructive to compare the present results with the mean-field approximation in $U$. In the mean-field approximation, there are two types of the solutions: the magnetic and nonmagnetic ones. The magnetic solution appears only in the strong Coulomb interaction case, as known in the Anderson model \cite{Anderson}. Although the magnetic solution does not describe the correct physics at zero temperature, we can infer the enhancement of the magnetic correlations at the QD via its existence. Hence, the mean-field analysis highlights the importance of the correlation effects, as discussed in the Anderson model out of equilibrium \cite{Komnik}.

Figure \ref{fig:mag} shows the local magnetization $m=(\ave{n_{\uparrow}}-\ave{n_{\downarrow}})/2$ at the QD as a function of $V$ and $\Gamma_{S}$. In the case of $0.25 \lesssim \Gamma_{S}/U \lesssim 0.3$, it is noteworthy that the magnetic solution exists only at a finite bias voltage; the bias voltage induces the local moment at the QD. With further increase in $V$, however, the magnetic solution disappears again. Such a reentrant behavior is not found in the nonequilibrium N-QD-N system for $\epsilon_{d}/U=-0.5$~\cite{Komnik}. Accordingly, we attribute the reentrant behavior to the competition/cooperation of the magnetic and superconducting correlations at a finite bias voltage. The reentrant behavior of the magnetic boundary clearly indicates that the magnetic correlation is enhanced by the bias voltage. Of course, the magnetic state is an artifact of the approximation and should be replaced by the Kondo singlet state, but the reentrant behavior of the boundary is consistent with the enhancement of the Kondo correlation deduced from the MPT calculation.

\begin{figure}
\includegraphics[width=8cm,clip]{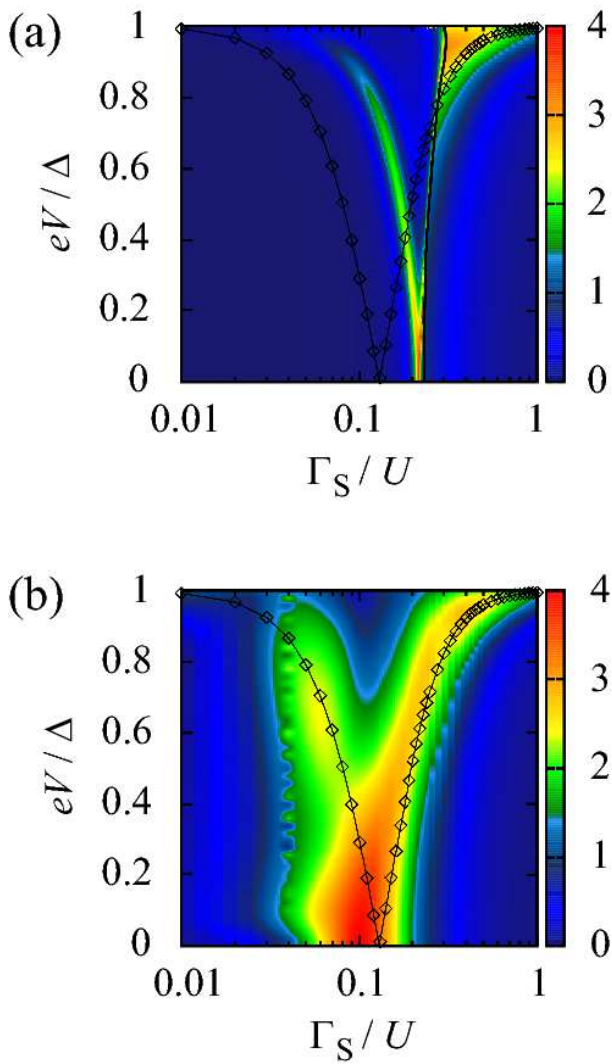}
\caption{\label{fig:cnd_cmpMF} (Color online) (a) Mean-field approximation results of the differential conductance: $\Gamma_{N}/U=0.05$, $\epsilon_{d}/U=-0.5$, $\Delta/U=0.025$ and $k_{B}T/\Gamma_{N}=0.01$. The energy of the Andreev bound states, $\omega_{b}/\Delta$, as a function of $\epsilon_{d}/U$ is denoted by diamonds. The black line indicates the boundary of the magnetic solution. (b) MPT results of the differential conductance. The parameters are same as those in (a).}
\end{figure}

We also calculate the differential conductance, $dI/dV$, with the mean-field approximation. Figure~\ref{fig:cnd_cmpMF}(a) shows the false-color scale representation of $dI/dV$ as functions of $eV/\Delta$ and the logarithm of $\Gamma_{S}/U$. The black solid line indicates the magnetic boundary, and the diamond denotes the energy of the Andreev bound states. In the region where the magnetic solution exists, the ridge position of $dI/dV$ features a curve similar to the energy of the Andreev bound states. On the other hand, in the large $\Gamma_{S}$ region where the non-magnetic solution only exists, the ridge is located around the gap edge and coincides with the Andreev bound states. As a result, the mean-field solutions still capture the essential nature of the Andreev bound states, except around the boundary line. At the boundary, the conductance changes in a discontinuous manner, implying that the physics related to the Andreev bound states can be qualitatively understood without quantum fluctuations~\cite{KoertingV10ntv}.

On the other hand, quantum fluctuations induce the intriguing phenomena a la Kondo, which cannot be described at the mean-field level in the region where the magnetic and superconducting correlations compete with each other. Figure~\ref{fig:cnd_cmpMF}(b) shows the MPT results of $dI/dV$. In addition to the good correspondence between the ridge of $dI/dV$ and the energy of the Andreev bound states, the anomalous enhancement occurs around the transition point  of the ground state in the QD-S system, which is caused by the Kondo-assisted Andreev reflection.  The quantum fluctuation effects are significant at a finite bias voltage in the vicinity of the transition point; The Kondo-assisted Andreev reflection, as well as the pinning effect of the Andreev resonances discussed in the previous section, is seen only at a finite bias voltage and smeared away from the transition point. Applying the bias voltage resembles increasing temperature concerning the enhanced weight of the excited states. 

\subsection{Nonequeliburium transport for generic ($\epsilon_{d}$, $V$) cases: comparison with experiments}

In the remainder of the section, we discuss the differential conductance as functions of the energy level at the QD, $\epsilon_{d}$, and the bias voltage, $V$. Note that  $\epsilon_{d}$ can be easily controlled by the gate voltage in actual experiments. We compare the analysis done here with the recent experiments qualitatively in good agreement.~\cite{DeaconRS10tso, DeaconRS10kat}

\subsubsection{Transport via Andreev bound states}

Let us first focus on the simple cases where the Kondo-assisted Andreev transport is not so important, for which we can highlight the crossover behavior in the transport due to the Andreev resonances.

\begin{figure*}
\includegraphics[width=17cm,clip]{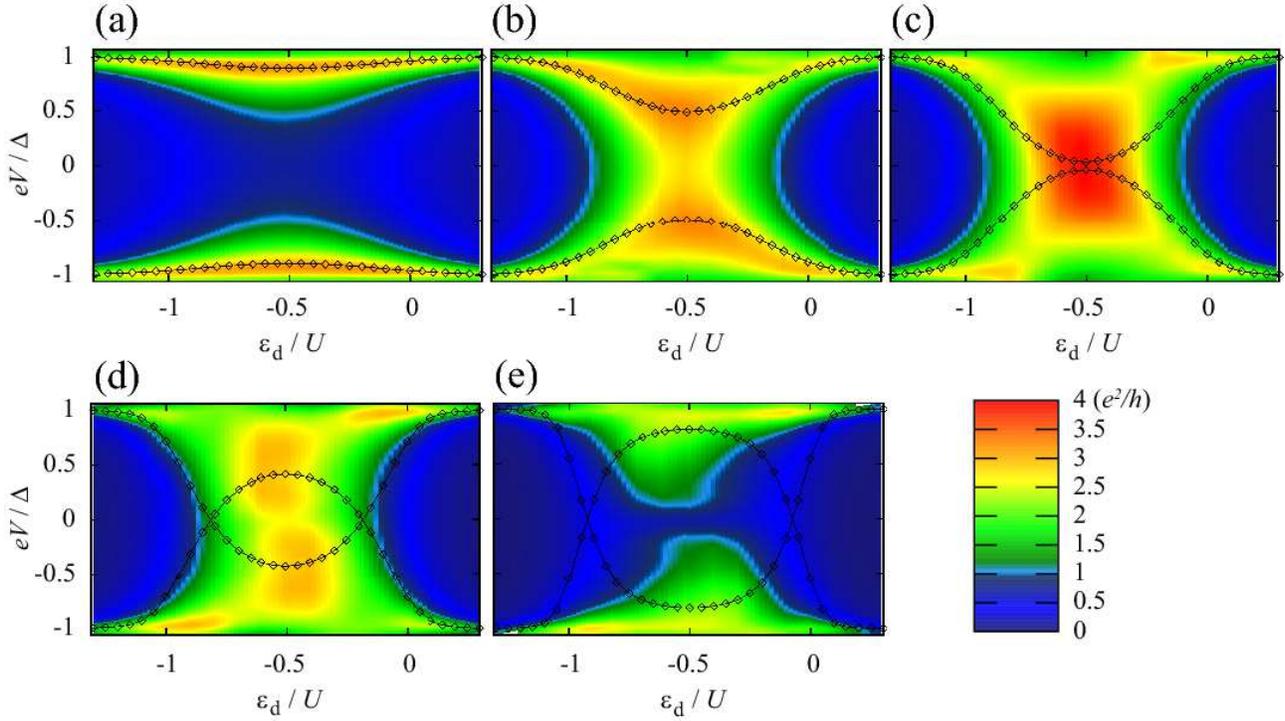}
\caption{\label{fig:cnd_EV} (Color online) False color-scale representation of the differential conductance as functions of $V$ and $\epsilon_{d}$ for $\Gamma_{N}/U=0.1$, $\Delta/U=0.05$ and $k_{B}T/\Gamma_{N}=0.01$: (a) $\Gamma_{S}/U=0.5$, (b) $\Gamma_{S}/U=0.25$, (c) $\Gamma_{S}/U=0.16$, (d) $\Gamma_{S}/U=0.1$, (e) $\Gamma_{S}/U=0.05$. The energies of the Andreev bound states, $\omega_{b}/\Delta$,  are also plotted.}
\end{figure*}

Figure \ref{fig:cnd_EV} is the false color-scale plot of $dI/dV$ as a function of $\epsilon_{d}$ for several values of $\Gamma_{S}$. We also show the $\epsilon_{d}$ dependence of the energy of Andreev bound states, which is calculated for $\Gamma_{N}=T=0$ with using the NRG method~\cite{BauerJ07spo}. In Fig. \ref{fig:cnd_EV}(a) and (b), the system is in the S-lead dominant coupling regime. It is clearly seen that the value of the bias voltage where  $dI/dV$ takes a peak value  coincides approximately with the energy of the Andreev bound states. The peak ridges correspond to Peak C defined in the previous subsection. For $\Gamma_{S}/U=0.16$, the system shows a crossover around $\epsilon_{d}/U=-0.5$ and the conductance has the maximum value, $4e^2/h$ (center of the figure \ref{fig:cnd_EV}(c)). Note that the energy of Andreev bound states does not touch the zero axis since $\Gamma_{S}/U=0.16$ characterizing the crossover is slightly different from the exact transition point. However, the peak positions still approximately correspond to the energy of the Andreev bound states.

With decreasing $\Gamma_{S}$, the system enters the N-lead dominant coupling regime as shown in Fig. \ref{fig:cnd_EV}(d). In the case of $\Gamma_{N}=T=0$ with the other parameters same as in (d), the magnetic doublet becomes the ground state around $\epsilon_{d}/U=-0.5$, and a transition occurs, where the softening of the Andreev bound states occurs around $\epsilon_{d}/U \simeq -0.5 \pm 0.32$. After the transition, the Andreev bound states move toward the gap edges.  For finite $\Gamma_{N}$, the magnetic doublet is changed to the Kondo singlet, which is realized around the center of the figure. Regarding $dI/dV$, the peaks corresponding to the Andreev bound states are not so clearly seen as in the S-lead dominant coupling regime. This is because $\Gamma_{N}$ is not small in comparison with $\Gamma_{S}$. The peaks around the symmetric point are especially indistinct because there are two kind of peaks of $dI/dV$: Peak A originating from the Kondo-assisted Andreev reflection and Peak B corresponding to the Andreev bound states.   With further decreasing  $\Gamma_{S}$, the bare value of the coupling of the  S-lead, $\Gamma_{S}$, becomes smaller than that of the N-lead, $\Gamma_{N}$. For  $\Gamma_{N}>\Gamma_{S}$, these peaks almost disappear in the subgap voltage. However, there is a remnant of the Kondo-assisted Andreev reflection, which makes a dip at $V=0$ in Fig. \ref{fig:cnd_EV}(e). For $\Gamma_{S}\to 0$, the remnant is also diminished and there are only peaks at the gap edges as in the noninteracting N-lead dominant coupling regime. 

As a result, both for $\Gamma_{N}/\Gamma_{S} \to 0$ and $\infty$, the subgap conductance, except at the gap edges, is completely suppressed for any values of $\epsilon_{d}$. A wide variety of patterns of the differential conductance result from the competition between the Coulomb interaction and the superconducting correlations at the QD.  

It is to be noted here that  the pronounced gap-edge peaks in the cases of $\Gamma_{N} \ll \Gamma_{S}$ and $\Gamma_{N} \gg \Gamma_{S}$, and the prominent peaks corresponding to the Andreev bound states in the S-lead dominant regime are qualitatively in agreement with the recent experimental results~\cite{DeaconRS10tso, DeaconRS10kat}. In the experiment~\cite{DeaconRS10tso, DeaconRS10kat}, the softening of the Andreev bound states in the N-lead dominant coupling regime is also observed in the distinctive peaks in the $dI/dV$ measurement. However, we do not find such a distinctly separated peak in the case $\Gamma_{N} \ge \Gamma_{S}.$  We will discuss the visibility of the peaks separately below.

\begin{figure*}
\includegraphics[width=17cm,clip]{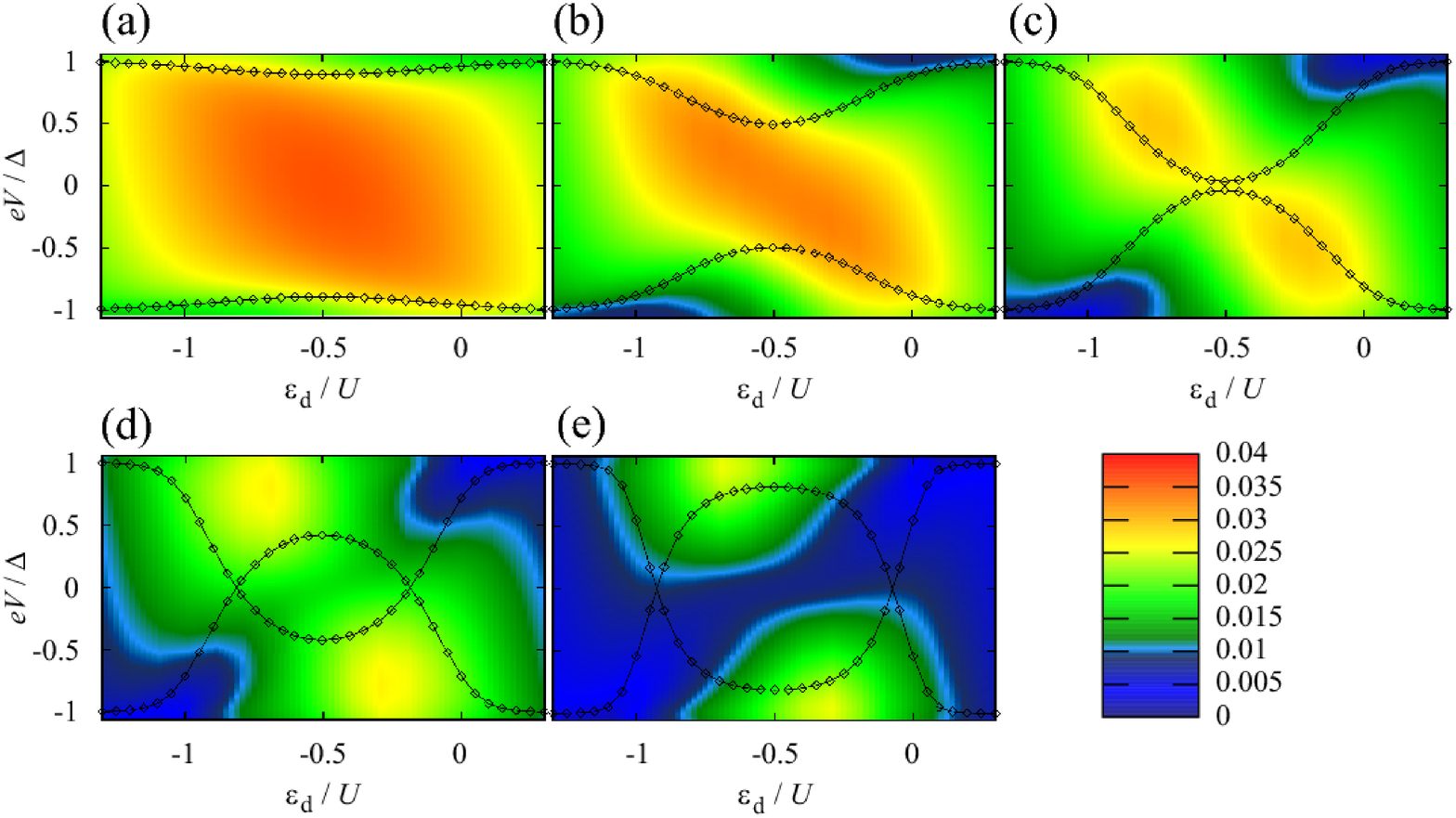}
\caption{\label{fig:Psi_EV} (Color online) False color-scale representation of $|\ave{d_{\downarrow}d_{\uparrow}}|$ as functions of $V$ and $\epsilon_{d}$ for $\Gamma_{N}/U=0.1$, $\Delta/U=0.05$ and $k_{B}T/\Gamma_{N}=0.01$: (a) $\Gamma_{S}/U=0.5$, (b) $\Gamma_{S}/U=0.25$, (c) $\Gamma_{S}/U=0.16$, (d) $\Gamma_{S}/U=0.1$, (e) $\Gamma_{S}/U=0.05$. The energies of the Andreev bound states, $\omega_{b}/\Delta$, as a function of $\epsilon_{d}/U$ are also plotted.}
\end{figure*}

\begin{figure}
\includegraphics[width=8cm,clip]{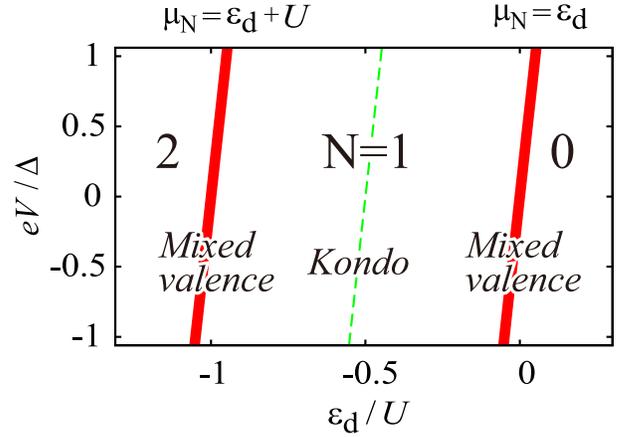}
\caption{\label{fig:phase_Gs0} (Color online) Schematic phase diagram of the N-QD-S system for $\Gamma_{S}=0$ and large $U$.}
\end{figure}

We also calculate the ($\epsilon_{d}$, $V$) dependence of  $|\ave{d_{\downarrow}d_{\uparrow}}|$ as shown in Fig. \ref{fig:Psi_EV}. For $\Gamma_{S}/U=0.5$, it is seen that $|\ave{d_{\downarrow}d_{\uparrow}}|$ is large around the center of the figure and decreases with increasing $\epsilon_{d}$ since the dominant SC-singlet state, which consists of the superposition of doubly-occupied and empty states, is weakened in the empty or doubly-occupied region. For finite $V$, $|\ave{d_{\downarrow}d_{\uparrow}}|$ decreases rapidly around the voltage  corresponding  to the energy of the Andreev bound states of a doublet character. For small $\Gamma_{S}$, the peak of $|\ave{d_{\downarrow}d_{\uparrow}}|$ is divided into two which are located around the voltage corresponding to the energy of the Andreev bound states of a SC-singlet character.  In spite that the electron and hole components of Andreev bound states have the same energy, there is asymmetry in $dI/dV$ and $|\ave{d_{\downarrow}d_{\uparrow}}|$ as a function of $V$. For instance, $|\ave{d_{\downarrow}d_{\uparrow}}|$ is strongly suppressed in the right top, in comparison with the one in the right bottom in Fig. \ref{fig:Psi_EV}. Moreover, the asymmetric feature becomes prominent with decreasing $\Gamma_{S}$. In order to explain how the asymmetry emerges, let us consider the $\Gamma_{S} \to 0$ limit. Figure \ref{fig:phase_Gs0} shows the schematic phase diagram of the N-QD system with strong $U$. The Kondo correlation is dominant for $\epsilon_{d}/U=-0.5$ and $V=0$, and is weakened with increasing $|\epsilon_{d}+0.5|$. Note that $V$ just shifts the chemical potential of the N-lead, so that only one of the two parameters, $eV$ and $\epsilon_{d}$, becomes relevant; the system stays in the same state along $\mu_{N}=\epsilon_{d}$ line in the figure. For $-0.5 + \Delta/U <\epsilon_{d}/U$, therefore, the Kondo correlation is enhanced if we fix $\epsilon_{d}$ and increase $V$. This enhancement of the Kondo correlation would occur in the N-QD-S system with small $\Gamma_{S}$, giving rise to the asymmetric patterns.

\subsubsection{How to observe Kondo-assisted Andreev transport }

The subgap peak structure in the differential conductance in the N-lead dominant coupling regime is a bit more complicated than that in the S-lead dominant coupling regime because both of the Kondo-assisted Andreev reflection and the Andreev bound states could affect the conductance profile as seen in Fig. \ref{fig:cnd_EV}. Therefore how we can observe the conductance peaks in the N-lead dominant coupling regime depends sensitively on the ratio of the bare coupling strengths. We find that if the N-lead dominant system is in the condition $\Gamma_{N}/\Gamma_{S}<1$, that is if $\widetilde{\Gamma}_{N}/\widetilde{\Gamma}_{S}>1$ and $\Gamma_{N}/\Gamma_{S}<1$, the conductance peaks become prominent in the subgap voltage.  We show some results below in the case satisfying this specific condition.

\begin{figure}
\includegraphics[width=8cm,clip]{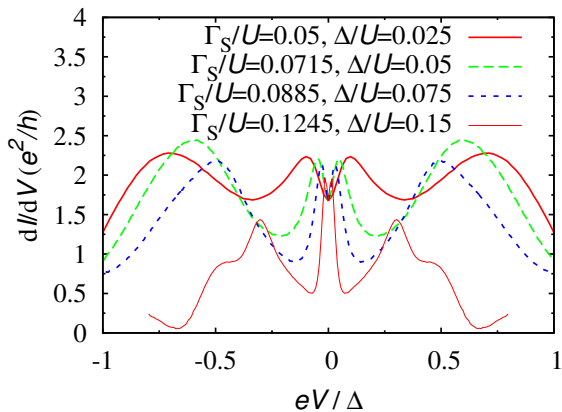}
\caption{\label{fig:cnd_atomiclimit} (Color online) Differential conductance as a function of $V$ for several values of ($\Gamma_{S}$, $\Delta$): $\Gamma_{N}/U=0.05$, $\epsilon_{d}/U=-0.5$, and $k_{B}T/\Gamma_{N}=0.01$.}
\end{figure}

Figure \ref{fig:cnd_atomiclimit} shows the differential conductance for several values of ($\Gamma$, $\Delta$) and $\epsilon_{d}/U=-0.5$. Note that the system is in the N-lead dominant coupling regime for any choices of  ($\Gamma$, $\Delta$) (see also the phase diagram of Fig. \ref{fig:phasediagram}). The two peaks in the vicinity of $V=0$ are related to the Kondo-assisted Andreev reflection and the other two at $eV/\Delta \simeq \pm 0.5$  originate from the Andreev bound states. These four peaks become sharper and more prominent from the top to bottom lines. In particular, the Kondo-like peaks approach the zero bias voltage since the peaks are located at $eV \simeq \pm \widetilde{\Gamma}_{N}$ and $\widetilde{\Gamma}_{N}$ decreases from top to bottom. Therefore, we conclude that for $U \gg \Gamma_{S}, \Delta \gg \Gamma_{N}$, only a single peak around $V=0$, instead of the two peaks at finite bias voltages, may be observed in real experiments.

\begin{figure*}
\includegraphics[width=17cm,clip]{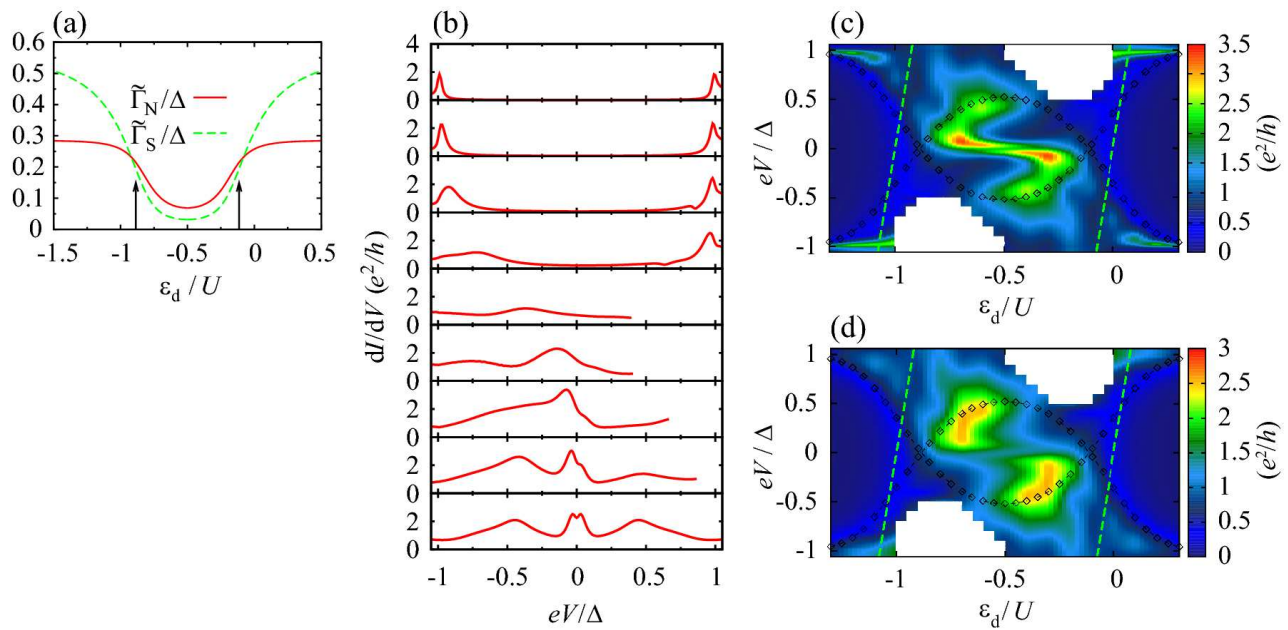}
\caption{\label{fig:cnd_EV_Kondo} (Color online) (a) Renormalized coupling strengths as a function of $\epsilon_{d}$ for $\Gamma_{N}/U=0.05$, $\Gamma_{S}/U=0.1$, $\Delta/U=0.075$ and $T=V=0$. (b) Differential conductance as a function of the bias voltage for several values of $\epsilon_{d}/U$ and $k_{B}T/\Gamma_{N}=0.01$. From bottom to top, $\epsilon_{d}/U=-0.5$, $-0.4$, $-0.3$, $-0.2$, $-0.1$, $0$, $0.1$, $0.2$ and $0.3$. The other parameters are the same as in (a). (c) False color-scale representation of the differential conductance as a function of $V$ and $\epsilon_{d}$. The parameters are the same as in (b). The energies of the Andreev bound states, $\omega_{b}/\Delta$, are denoted as diamonds. The bound states exist for $\Gamma_{N}=T=0$. The dashed lines indicate the resonant conditions; $\mu_{N}=\epsilon_{d}$ and $\mu_{N}=\epsilon_{d}+U$. (d) At $k_{B}T/\Gamma_{N}=0.1$ which is nearly equal to that of $\widetilde{\Gamma}_{N}/\Gamma_{N}$, the central Kondo-enhanced Andreev ridges are suppressed.}
\end{figure*}

We next look at the $\epsilon_{d}$ dependence of the N-QD-S system for $\Gamma_{N}/U=0.05$, $\Gamma_{S}/U=0.1$ and $\Delta/U=0.075$ (Fig. \ref{fig:cnd_EV_Kondo}). In this case, $\widetilde{\Gamma}_{N}>\widetilde{\Gamma}_{S}$ is satisfied at the particle-hole symmetric point ($\epsilon_{d}/U=-0.5$) in addition to the condition of $\Gamma_{N}/\Gamma_{S}<1$, which leads to the crossover in the dominant couplings may occur away from the symmetric point as shown Fig. \ref{fig:cnd_EV_Kondo}(a). The crossover points are approximately coincide with the doublet-singlet transition points denoted by arrows for $\Gamma_{N}=0$. In the particle-hole symmetric case, $\widetilde{\Gamma}_{N}/ \Gamma_{N} \simeq 0.1$ characterizes the position and the temperature dependence of the Kondo-type peaks in $dI/dV$. For $k_{B}T<\widetilde{\Gamma}_{N}$, both of the Kondo-assisted Andreev reflection and the Andreev bound states at the QD contribute to the nonlinear electron transport. Figure \ref{fig:cnd_EV_Kondo}(b) shows the $\epsilon_{d}$ dependence of $dI/dV$ at $k_{B}T/\Gamma_{N}=0.01$. Regarding the central two peaks related to the Kondo effect, one of the peaks increases its height away from $\epsilon_{d}/U=-0.5$, whereas the other disappears.
 We note that the curve of $dI/dV$ ends at certain voltages for $\epsilon_{d}/U=-0.4$, $0.3$, $0.2$ and $-0.1$ since we cannot get the convergent solution in the framework of MPT. Therefore, the movement of the one of the two peaks owing to the Andreev bound states is not clear in this figure, unfortunately. Nevertheless, we can see that the other peak moves to lower bias voltages and merges into the single peak with the prominent peak related to the Kondo effect. With further increasing $\epsilon_{d}$, the merged peak moves toward the gap edge and there is no peak in the subgap voltage.

A quantitative comparison of the peak positions with the energy of the Andreev bound state, $\omega_{b}/\Delta$, is shown in Fig. \ref{fig:cnd_EV_Kondo}(c). Softening of the Andreev bound states clearly emerges in the peak structure of the differential conductance. The softening reflects the transition of the ground states in the case that $\Gamma_{N}=0$. For $-0.75 \lesssim \epsilon_{d}/U \lesssim -0.25$, there are two Kondo-assisted Andreev reflection ridges around $V=0$. The two ridges are separated around $\epsilon_{d}/U-0.5$. In the case of $\epsilon_{d}/U \lesssim -0.75$ or $-0.25 \gtrsim \epsilon_{d}/U $, the Kondo-type ridges are not observed since the system is away from the Kondo regime, and there appear only the gap-edge peaks related to the Andreev bound states. The dashed lines in the figure denote the conditions where the level of the QD and the chemical potential of the N-lead satisfy the equations; $\mu_{N}=\epsilon_{d}+U$ and $\mu_{N}=\epsilon_{d}$. In spite that the Andreev bound states have the symmetric energy spectrum, the large asymmetry is found in $dI/dV$ as a function of $V$, except for $\epsilon_{d}/U \neq -0.5$. The asymmetry reflects the fact that the Kondo correlations appear differently depending on the sign of the bias voltage, as discussed in Fig. \ref{fig:phase_Gs0}.

It is remarkable that the overall features of the differential conductance are consistent with those in the recent experiment~\cite{DeaconRS10kat}, including the observation of the fingerprints of two kinds of peaks; Kondo-type peaks at $eV/\Delta \simeq 0$ and the peaks due to the Andreev bound states. There still seems to be a small discrepancy between the theory and the experiment~\cite{DeaconRS10kat}. In our theory, the Kondo-type ridges are separated at the particle-hole symmetric point, while a single Kondo ridge is observed, instead of the two ridges, in the experiment. We believe that the single ridge is a consequence of the special condition, $U \gg \Gamma_{S}, \Delta \gg \Gamma_{N}$, used in the experiment. In this case, the two Kondo-type peaks for $\epsilon_{d}/U=-0.5$ would be located near $V=0$ and overlap with each other, as discussed in Fig. \ref{fig:cnd_atomiclimit}. Hence, the two Kondo ridges could be also connected at the particle-hole symmetric point in such a special condition.

Finally, some comment are in order for the temperature dependence. In Fig. \ref{fig:cnd_EV_Kondo}(d), we present $dI/dV$ at a higher temperature $k_{B}T_{K}/\Gamma_{N}=0.1$, which is comparable to $\widetilde{\Gamma}_{N}$. The Kondo-type ridges are smeared with increasing temperature, but the peaks due to the Andreev bound states show little change. Therefore, at higher temperatures, there would be only the crossover behavior of the fingerprint of the Andreev bound states in the $dI/dV$ measurements.

\section{Summary}

In this paper, we have theoretically investigated the nonequilibrium electron transport through a quantum dot coupled to the normal and superconducting leads with particular emphasis on the interplay between the Kondo effect and the superconducting correlations. For this purpose, we have developed the modified second order perturbation theory in Keldysh-Nambu formalism under nonequilibrium steady-state conditions. We have confirmed that this method is indeed efficient for analyzing the nonequilibrium electron transport in the present system.

It has been shown that the renormalized couplings between the leads and the dot in the equilibrium states are the key quantities that correctly describe nonequilibrium transport properties. In particular, the enhancement of the Andreev transport occurs via a Kondo resonance at a finite bias voltage, giving rise to an anomalous peak structure in the differential conductance, whose position is determined solely by the above-mentioned renormalized parameters. This peak formation  is a remarkable example of phenomena that are indeed caused by the interplay between the Kondo and  superconducting correlations. A pinning effect of the Andreev resonances to the Fermi level of the normal lead and its counter level also evidences the interplay of the above two types of correlations. Moreover, it has been shown that the energy levels of the Andreev bound states give rise to an additional peak structure in the differential conductance in the strongly correlated N-QD-S system. 

We have demonstrated that the above characteristic features of nonequilibrium differential conductance obtained from our calculation are qualitatively in agreement with those observed in the recent experiments~\cite{DeaconRS10kat,DeaconRS10tso}. Finally we note that there still exists a small discrepancy between the theory and the experiments; the Kondo-type ridges are separated in our theory (Fig. \ref{fig:cnd_EV_Kondo}), while a single Kondo ridge is observed experimentally.~\cite{DeaconRS10kat} We believe that the single ridge is a consequence of a specific condition employed in the experiments, $U \gg \Gamma_{S}, \Delta \gg \Gamma_{N}$ and that the clearly-separated Kondo ridges predicted in this paper will be observed if the proper conditions for the system parameters are prepared experimentally.

\begin{acknowledgments}
We would like to thank A.~Oguri, R.~S.~Deacon, M.~Imada and J.~Bauer for fruitful discussions. This work was supported by KAKENHI (Nos. 21740232, 20104010), the Grant-in-Aid for the Global COE Programs ``The Next Generation of Physics, Spun from Universality and Emergence'' from MEXT of Japan, and JSPS through its FIRST ProgramD Y. ~Yamada is supported by JSPS Research Fellowships for Young Scientists, and Y.~Tanaka is supported by Special Postdoctoral Researchers Program of RIKEN.

\end{acknowledgments}


\end{document}